\documentclass[prb,twocolumn,showpacs,superscriptaddress]{revtex4}
\usepackage[english]{babel}
\usepackage[dvips]{graphicx}
\usepackage{color}
\usepackage{latexsym}
\usepackage{amsmath}
\usepackage{amsthm}
\usepackage{amsfonts}
\usepackage{amssymb}

\begin{document}

\title{Quantum phase-slips in Josephson junction rings}

\author{G.\ Rastelli}
\affiliation{Universit{\'e} Grenoble 1/CNRS, LPMMC UMR 5493, B.P. 166, 38042 Grenoble, France}
\affiliation{Fachbereich Physik, Universit{\"a}t Konstanz, D-78457 Konstanz, Germany}

\author{I.\ M.\ Pop}
\affiliation{Institut N{\'e}el, CNRS, and Universit{\'e} Joseph Fourier, B.P. 166, 38042 Grenoble, France}
\affiliation{Department of Applied Physics, Yale University, New Haven, Connecticut 06520, USA}

\author{F.W.J.\ Hekking}
\affiliation{Universit{\'e} Grenoble 1/CNRS, LPMMC UMR 5493, B.P. 166, 38042 Grenoble, France}

\begin{abstract}
We study quantum phase-slip (QPS) processes in a superconducting ring
containing $N$ Josephson junctions and threaded by an external static
magnetic flux $\Phi_B$.
In a such system, a QPS consists of a quantum tunneling event connecting
two distinct classical states of the phases with different persistent currents
[K.\ A.\ Matveev et al., Phys.\ Rev.\ Lett.\ {\bf 89}, 096802 (2002)].
When the Josephson coupling energy $E_J$ of the junctions is larger than the charging
energy $E_C=e^2/2C$ where $C$ is the junction capacitance, the quantum amplitude
for the QPS process is exponentially small in the ratio $E_J/E_C$.
At given magnetic flux each QPS can be described as the tunneling of
the phase difference of a single junction of almost $2 \pi$, accompanied by a
small harmonic displacement of the phase difference of the other $N-1$ junctions.
As a consequence, the total QPS amplitude $\nu_\mathrm{ring}$ is a global property of the
ring.
Here we study the dependence of $\nu_\mathrm{ring}$ on the ring size $N$, taking into
account the effect of a finite capacitance $C_0$ to ground which leads to the appearance
of low-frequency dispersive modes.
Josephson and charging effects compete and lead to a non-monotonic dependence of
the ring's critical current on $N$.
For $N \rightarrow \infty$, the system converges either towards a superconducting
or an insulating state, depending on the ratio between the charging energy $E_0=e^2/2C_0$
and the Josephson coupling energy $E_J$.
\end{abstract}


\date{\today}

\maketitle

\section{Introduction}
One-dimensional Josephson junction chains (1D JJ chains) have received considerable
interest recently.
Their use has been proposed for the realization of a qubit topologically protected
against decoherence,\cite{Ioffe:2002Nature,Ioffe:2002,Douçot:2002,Douçot:2003,Douçot:2005,Gladchenko:2009}
for the realization of a tunable parametric amplifier in narrow frequency
ranges,\cite{Castellanos-Beltrana:2007,Castellanos-Beltrana:2008}
for the realization of a fundamental current standard
in quantum metrology,\cite{GuichardHekking:2010,Flowers:2004}
and for designing controlled inductive electromagnetic environments
in quantum circuitry.\cite{Manucharyan:2009,Haviland:2011}

\begin{figure}[thbp]
\includegraphics[scale=0.14,angle=0.]{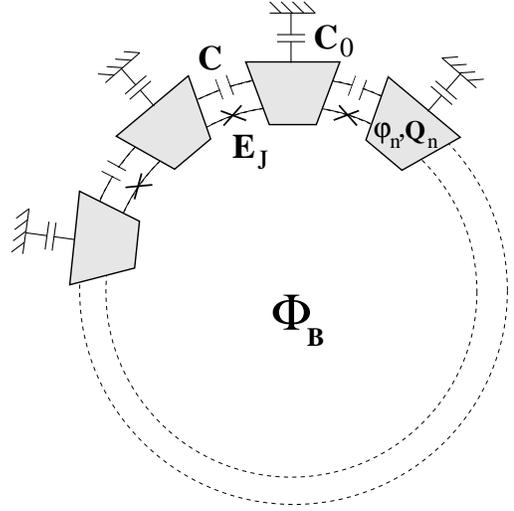}
\caption{
Schematic representation of a superconducting ring
threaded by a magnetic flux $\Phi_B$ and containing $N$ identical Josephson junctions
with a mutual capacitance $C$ and a local ground capacitance $C_0$. $E_J$ is the
energy scale for the Josephson coupling. The variables $(\varphi_n,Q_n)$ are, respectively,
the condensate phase and the excess charge of the $n$th superconducting island.}
\label{fig:scheme_array}
\end{figure}

Homogeneous JJ chains of infinite length have been studied theoretically in the
past.\cite{BradleyDoniach:1984,Korshunov:1986,Korshunov:1989,Fazio:2001,Sarkar:2007,Sarkar:2009}
Such chains consist of superconducting islands, separated by Josephson tunnel junctions.
In this paper we consider JJ chains arranged in a closed geometry, Fig.~\ref{fig:scheme_array}.
The electrostatic interaction between the metallic islands is modeled by a neighboring
capacitance $C$ and by a local ground  capacitance $C_0$, with $E_C=e^2/2C$ and $E_0=e^2/2C_0$
the corresponding charging energies.
Each Josephson junction can sustain a maximum supercurrent $I_J=2eE_J/\hbar$;
this defines the Josephson coupling energy $E_J$.
%
%
%%%%%%%%%%%%%%%%%%%%%%%%%%%%%%%%%%%%%%%%%%%%%%%%%%%%%%%%%%%%%%%%%%%%%%%
%
%
%   FIG.1
%
%

Previous theoretical studies\cite{BradleyDoniach:1984,Korshunov:1986,Korshunov:1989} predicted a
superconductor-insulator phase transition when the ratio between the Josephson energy
$E_J$ and the characteristic charging energy is reduced below a critical value.
Bradley and Doniach studied this phase transition for infinite JJ chains and for the two
extreme opposite cases when one of the two capacitances is vanishing
($C_0=0$ or $C=0$).\cite{BradleyDoniach:1984}
Korshunov investigated the general case for arbitrary ratio
$C/C_0$.\cite{Korshunov:1986,Korshunov:1989}
He found that the critical value $E^{(c)}_J$ of the Josephson energy at which
the system undergoes the phase transition equals to $E^{(c)}_J = E_0 f(C/C_0)$ where
$f(x)$ is a smooth and regular function of order one.
In particular, for the case $C_0=0$, we have $E_0=\infty=E^{(c)}_J$ and
the system is an insulator for {\em any value of $E_J$} in agreement with
the result of Bradley and Doniach.\cite{BradleyDoniach:1984}
Subsequently, experimental studies of the finite temperature behavior of the residual
resistance in long one-dimensional chains of SQUIDs reported a phase transition when
reducing the Josephson
energy.\cite{Chow:1998,Haviland:2000,Kuo:2001,Miyazaki:2002,Takahide:2006}

The theoretical results reviewed so far were obtained in the thermodynamic limit $N\to\infty$.
A first attempt to go beyond this limit was undertaken by Matveev, Glazman and
Larkin~\cite{MateevLarkinGlazman:2002} who studied quantum phase-slip (QPS) processes
in a superconducting nanoring containing a large, but finite number of Josephson junctions.
Here a QPS consists of a quantum tunneling between two distinct classical states
of the phases with different persistent currents circulating in the ring at given
magnetic flux.~\cite{MateevLarkinGlazman:2002}
This is a collective process which can be described as the tunneling of the
phase difference of a single junction by {\em almost} $2 \pi$, accompanied by a small
harmonic displacement of the phase difference of the other $N-1$ junctions
(see Ref.~\onlinecite{MateevLarkinGlazman:2002} and the explanation in Sec.~\ref{subsec:SPSap}).
Quantum tunneling is possible due to the finite junction capacitance,which plays the role of inertia.

Matveev et al. predicted a strong reduction of the maximum critical current
sustained by the ring with increasing ring size $N$ due to QPS processes.
Recent experiments on flux-biased rings containing a few Josephson
junctions\cite{Pop:2010} indeed reported a remarkable suppression of the maximum
supercurrent as $E_J/E_C$ decreases, in agreement with the findings of
Ref.~\onlinecite{MateevLarkinGlazman:2002}.
In these devices, the effects of the capacitance to ground could be neglected
since the ring's circumference was much smaller than the screening length $\lambda$ of the
system, given by $\lambda=\pi \sqrt{C/C_0}$.
However, it is expected that, for JJ rings of intermediate circumference $N \agt \lambda$,
the effects of the capacitance  to ground can no longer be ignored.

In this paper, we study a JJ ring of finite circumference and threaded by an external
magnetic flux $\Phi_B$, Fig.~\ref{fig:scheme_array}.
Specifically, we consider properties of the flux-dependent thermodynamic persistent current.
We go beyond the previous work of Matveev et al.\cite{MateevLarkinGlazman:2002}
and we take into account the collective nature of a QPS as well as the ground capacitance $C_0$
for calculating the QPS amplitude.
We show that the interplay between the finite value of the ratio $C_0/C$
and finite size effects gives rise to a non-monotonic dependence of the
low-energy properties on $N$.
We systematically discuss this interplay as well as its consequences for the QPS
amplitude for flux-biased rings with arbitrary number $N \agt 5$. For shorter lengths, a
detailed numerical analysis was realized in~\onlinecite{Orlando:1999}.

We focus on the limit where the Josephson coupling energy $E_J$ dominates over the charging
energies $E_C,E_0$, such that the amplitude for QPS to occur is exponentially
small in the ratio $E_J/E_C$.
This fact allows us to focus on the analysis of a single QPS event.
Once the QPS amplitude is known, one can calculate the ring's low-energy spectrum
as a function of the external flux $\Phi_B$  and hence obtain the maximum supercurrent
$I_\mathrm{max}$ that the ring can sustain.

\section{Qualitative discussion and main results}
\begin{figure}[hbtp]
\includegraphics[scale=0.15,angle=0.]{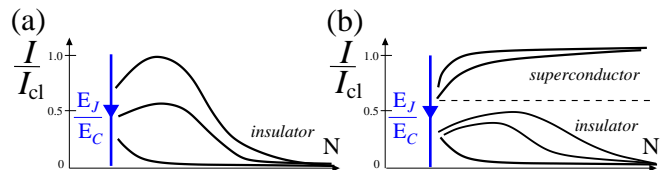}
\caption{Schematic behavior of the maximum supercurrent $I_\mathrm{max}$ in 1D JJ rings
as a function of  $N$, scaled to the classical value  $I_\mathrm{cl}$, for
different values of the ratio $E_J/E_C$ for (a) $C_0=0$ and (b) $C_0=C/2$
(see also Fig.~\ref{fig:I_ren_1} and Fig.~\ref{fig:I_ren} for details).}
\label{fig:scheme_results}
\end{figure}

Our main results are summarized in Fig.~\ref{fig:scheme_results}, where we show the
dependence of $I_\mathrm{max}$ -- scaled to the classical value
$I_\mathrm{cl}=\pi I_J/N$ found in the absence of QPS processes --
as a function of $N$ for two relevant situations: $C_0=0$ and $C_0=C/2$.
%
%
%
%%%%%%%%%%%%%%%%%%%%%%%%%%%%%%%%%%%%%%%%%%%%%%%%%%%%%%%%%%%%%%%%%%%%%%%
%
%
%   FIG.2
%
%

As we will discuss in detail below, a QPS  can be described as a $2\pi(1-1/N)$ winding of
the local phase-difference  occurring on one of the junctions, accompanied by a
simultaneous small (harmonic) adjustment of the phases of the other $N-1$
junctions.\cite{MateevLarkinGlazman:2002}
In a first approximation, the winding  of the phase on one of
the junctions can be characterized by the amplitude for the quantum tunneling between two
different minima of the Josephson potential.
It is given by\cite{Likharev:1985,MateevLarkinGlazman:2002,Catelani:2011,note:factor_2}
\begin{equation}
\label{eq:v_0}
\nu_0=\frac{4}{\sqrt{\pi}} {\left(8 E_J^3 E_C\right)}^{\frac{1}{4}} \exp\left(- \sqrt{8
\frac{E_J}{E_C}}\right) \, .
\end{equation}
The dynamics of the simultaneous small adjustment depends crucially on the
capacitance ratio $C_0/C$.

Consider first the case $C_0=0$, Fig.~\ref{fig:scheme_results}(a).
The other $N-1$ junctions form a bath of dispersionless harmonic oscillators,
all having the same (plasma) frequency $\omega_p={\left(8E_JE_C\right)}^{1/2}/\hbar$.
In order to satisfy the constraint imposed by the flux threading the ring at all times
during the QPS process, the phase differences for the $N-1$ other junctions
perform a small shift $\sim 1/N$.
This adjustment gives rise to finite-size corrections for intermediate ring
circumferences $N$ to the amplitude given by Eq.~(\ref{eq:v_0}) leading to a QPS
amplitude $\nu_0\rightarrow\nu(N)$.
Since any junction can act as a QPS center, the total QPS amplitude for
the ring is given by $\nu_\mathrm{ring}=N\nu(N)$.

In the limit $N\gg 1$, finite size effects vanish and
$\nu(N)$ converges to the constant  $\nu_0$ so that
$\nu_\mathrm{ring}$ increases with the length and
the maximum supercurrent $I_\mathrm{max}$ vanishes
exponentially,\cite{MateevLarkinGlazman:2002} see Fig.~\ref{fig:scheme_results}(a).
The system becomes a perfect insulator at $N=\infty$, in agreement with Refs.
\onlinecite{BradleyDoniach:1984,Korshunov:1986,Korshunov:1989}.
On the other hand, we find that the interplay between charging and Josephson
effects in finite systems leads to an {\em enhancement} of the effective QPS amplitude
$\nu(N)$ with {\em decreasing} ring circumference $N$, $\nu(N) \sim \nu_0 \exp[
{\left(E_J/E_C\right)}^{1/2}/N]$, thus {\em} reducing the maximum supercurrent
$I_\mathrm{max}$.
Consequently, the maximum supercurrent $I_\mathrm{max}$ shows
non-monotonic behavior as a function of the ring circumference $N$ for sufficiently large
values of the ratio $E_J/E_C$.

When the capacitance to ground $C_0$ is restored, the $N-1$ harmonic junctions interact
directly between them.
This leads to the appearance of an ensemble of $N-1$ dispersive electrodynamics modes at
frequencies below $\omega_p$, similar to the ones found in a standard LC-transmission
line.\cite{Masluk:2012}
The tunneling phase couples to these modes in much the same way as a quantum particle to
a harmonic bath in the Caldeira-Leggett model.\cite{Caldeira:1981}
In particular, the low-frequency modes with linear dispersion $\hbar \omega_k
\sim {\left(8E_JE_0\right)}^{1/2} \pi k/N$ give rise to a finite friction for the
QPS dynamics in the limit $N=\infty$.

At finite $N$, the coupling with the low-frequency modes strongly
affects the QPS amplitude.
Indeed, we find that $\nu(N) \sim
\nu_0 /N^{\alpha}$ for $N \gg 1$ where
$\alpha \propto {\left(E_J/E_0\right)}^{1/2}$.
Depending on the value of $\alpha$,
$\nu_\mathrm{ring} = N \nu(N) \sim N^{1-\alpha}$ either tends to zero, when $\alpha \gg
1$, or grows linearly, when $\alpha \ll 1$, indicating that the system either displays a
superconducting or an insulating behavior, as can be seen in Fig.~\ref{fig:scheme_results}(b).
This behavior is reminiscent of the dissipative phase transition\cite{Schmid:1983}
occurring in a single junction in an electromagnetic environment.\cite{Korshunov:1986,Korshunov:1989}
For intermediate ring sizes, finite size effects occur, yielding a non-monotonic behavior
of the maximum supercurrent in the insulating regime, similar to what we find for the
case $C_0=0$.

The paper is structured as follows.
In Sec.~\ref{sec:model} we recall the model for a flux-biased 1D JJ ring  as well as
the notion of QPS and the approximations involved to find the ring's flux-dependent
quantum ground state and hence the maximum supercurrent.
In Sec.~\ref{sec:1PS_approximation} we discuss the single QPS approximation and
we show how the system reduces to a model similar to that of Caldeira
and Leggett\cite{Caldeira:1981} where one single Josephson junction,
the center of the QPS, is coupled to $N-1$ harmonic oscillators.
The results for the specific case $C_0=0$ are shown in Sec.~\ref{sec:result_C0=0}
where we explain in detail the different finite-size corrections
on the QPS amplitude.
The effect of the finite ground capacitance $C_0>0$ and
the general results are discussed in Sec. \ref{sec:results_general}.
In the last Section \ref{sec:conclusions} we draw our conclusions.

\section{The model}
\label{sec:model}

\subsection{Hamiltonian}
\label{sec:model-Hamiltonian}
We consider a homogeneous ring of $N$ identical superconducting islands each coupled to
its nearest neighbor by Josephson tunnel junctions, see Fig.~\ref{fig:scheme_array}.
The ring is threaded by a magnetic flux $\Phi_B$.
We assume the superconducting gap $\Delta$ of the islands to be the
largest energy scale
involved in the problem. If $\Delta \gg \delta E$
where $\delta E$ is the average spacing of the electronic energy levels,
superconductivity is well established.
The islands should be metallic with large enough volume so that
the perturbative treatment of Cooper pair tunneling through the
contacting surfaces is justified.
We assume the absence of quasi-particle excitations at low temperature $T$ and
low voltage $k_BT, 2e\bar{V}\ll \Delta$, where $\bar{V}$ is the typical voltage across
the junctions.
Furthermore, the condition $\Delta \gg  E_C,E_0$ implies that the Josephson
coupling energy $E_J$ characterizing Cooper-pair-tunneling between islands
is independent of $E_C, E_0$.

We assume that the kinetic inductance, associated with the
kinetic energy of the Cooper pairs in each superconducting island,
is  negligible as compared to the Josephson inductance.\cite{note:charge-solitons,Hekking:1997}
We also assume that the  geometric inductance can be neglected
so that the current circulating in the ring does not generate any magnetic field.
The total flux $\Phi_B$ is thus only given  by the externally applied magnetic field.

The previous conditions  define the standard quantum-phase model for
a 1D-JJ  homogeneous  chain whose Hamiltonian reads\cite{Fazio:2001}
\begin{equation}
\label{eq:Hamiltonian}
H \!\!=\!\!
\frac{1}{2} \! \sum^{N-1}_{n,m=0} \!\!\! \hat{Q}_n  \bar{\bar{C}}^{-1}_{nm} \hat{Q}_m
-
E_J \!\!\sum^{N-1}_{n=0} \!\! \cos\left( \! \hat{\varphi}_{n+1} -\hat{\varphi}_{n}
+ \frac{2\pi\Phi_B}{N\Phi_0} \! \right)  \, ,
\end{equation}
where $\Phi_0$ is the flux quantum.
For each island, the BCS condensate phase $\hat{\varphi}_n$ and the excess
charge $\hat{Q}_n$ on the $n$th island represent the conjugate variables of the system
$[ \hat{\varphi}_n, \hat{Q}_n ]=2e i$.
$ \bar{\bar{C}}$ is the capacitance matrix with matrix elements
$\bar{\bar{C}}_{n,m}=(C_0+2C) \delta_{n,m} - C (\delta_{n+1,m} +
\delta_{n-1,m})$, with the index $n=-1$ corresponding to $N-1$ and
$n=0$ corresponding to $N$.
The relative phase-difference across the $n$th junction is
$\hat{\theta}_n = \hat{\varphi}_{n+1} -\hat{\varphi}_{n}$.
As the phases are compact variables, {\em i.e.} $\hat{\varphi}_N = \hat{\varphi}_0 + 2\pi
m$ where $m$ is an integer, we have the constraint on the phase-differences for Josephson
junctions in a ring\cite{Tinkham:1996}
\begin{equation}
\label{eq:constraint}
\sum^{N-1}_{n=0} \hat{\theta}_n = 2\pi m  \, .
\end{equation}
Note that the argument of each cosine in Eq.~(\ref{eq:Hamiltonian}) is
the gauge-invariant phase-difference across the corresponding junction.

From Eqs.~(\ref{eq:Hamiltonian}) and (\ref{eq:constraint}) we see that
the physical properties of the system depend periodically on the ratio
$\delta=2\pi \Phi_B/\Phi_0$.
In the steady-state regime, the dc-supercurrent flowing through the ring
is the same for all the junctions $\langle\hat{I}_n \rangle =I$ and can be related
to the derivative of the ground state energy $E_{GS}$ of the system with respect to $\delta$
\begin{equation}
\label{eq:I_delta}
I\left(\delta\right) = \frac{\partial  E_{GS}}{\partial \Phi_B} =
\left(\frac{2e}{\hbar}\right)  \frac{\partial  E_{GS}}{\partial \delta} \, .
\end{equation}
It is in general a difficult task to find the ground-state energy $E_{GS}(\delta)$ for
the flux-biased ring described by Hamiltonian (\ref{eq:Hamiltonian})
and (\ref{eq:constraint}).
An approximate solution can be found in the limit where the Josephson energy is larger
than the characteristic electrostatic energy $E_J \gg E_C, E_0$, which is the regime
discussed in this paper.

\subsection{Single QPS in JJ rings of finite circumference}
\label{subsec:SPSap}
To set the stage, let us first consider the classical limit, achieved by setting
$E_C=E_0=0$, so that the phases are well-defined classical variables.
The classical energy of the system reduces to
\begin{equation}
\label{eq:E_classical}
E_{cl} = - E_J \sum^{N-1}_{n=0} \cos\left(\theta_n  + \frac{\delta}{N} \right) \, .
\end{equation}
The energy Eq.~(\ref{eq:E_classical}) is invariant under a change by $2\pi$ of the phases
$\theta_n$.
In other words, the states $\theta_n$ and $\theta_n+2\pi k$ are equivalent ($k$ integer).
However, at fixed magnetic flux $\delta$, a given configuration of $\left\{ \theta_n \right\}$
corresponds to a real  physical state {\sl only if} the constraint Eq.~(\ref{eq:constraint}) is satisfied.
Therefore any distribution of the phases that violates Eq.~(\ref{eq:constraint}) is unphysical.

The classical states $\left| m \right>$  that minimize the energy Eq.~(\ref{eq:E_classical})
under the constraint (\ref{eq:constraint}) correspond to a uniform distribution of phase
differences $\theta_n = 2\pi m /N$.
They have energies
\begin{equation}
\label{eq:E_m}
E_{m} = - E_J N \cos \left( \frac{2\pi m + \delta}{N} \right)  \, ,
\end{equation}
with the condition ${\left. d^2 E_{cl}/d\theta_n^2 \right|}_{m} = -E_m > 0$ and
the index $-(N-1)/2<m<(N-1)/2$ ($N$ odd) or $-N/2<m<(N/2)-1$ ($N$ even).
These classical states $\left| m \right>$ are physically {\sl distinguishable} as they are
characterized by different persistent currents
\begin{equation}
\label{eq:I_m}
I_{m} = I_J \,  \sin \left( \frac{2\pi m + \delta}{N} \right)  \, .
\end{equation}
Away from the degeneracy points $\delta=0$ and $\delta=\pi$ they also have different
energies.
The classical ground state corresponds to an absolute minimum
\begin{equation}
\label{eq:E_GS_cl}
E^{(cl)}_{GS} \!\!= \! -E_J N \max_m \cos \left( \frac{2\pi m + \delta}{N} \right)
\!\! \simeq \!\! \frac{E_J}{2N}
\min_m {\left( 2\pi m + \delta \right)}^2  ,
\end{equation}
where the second, approximate equality is numerically accurate for sufficiently long
rings $(N \agt 5)$.
The corresponding supercurrent then has as a sawtooth-like dependence as a
function of $\delta$ with a maximum supercurrent given by
$I_\mathrm{cl} \simeq \pi I_J/N$.\cite{MateevLarkinGlazman:2002}

\begin{figure}[htbp]
\includegraphics[scale=0.25,angle=0.]{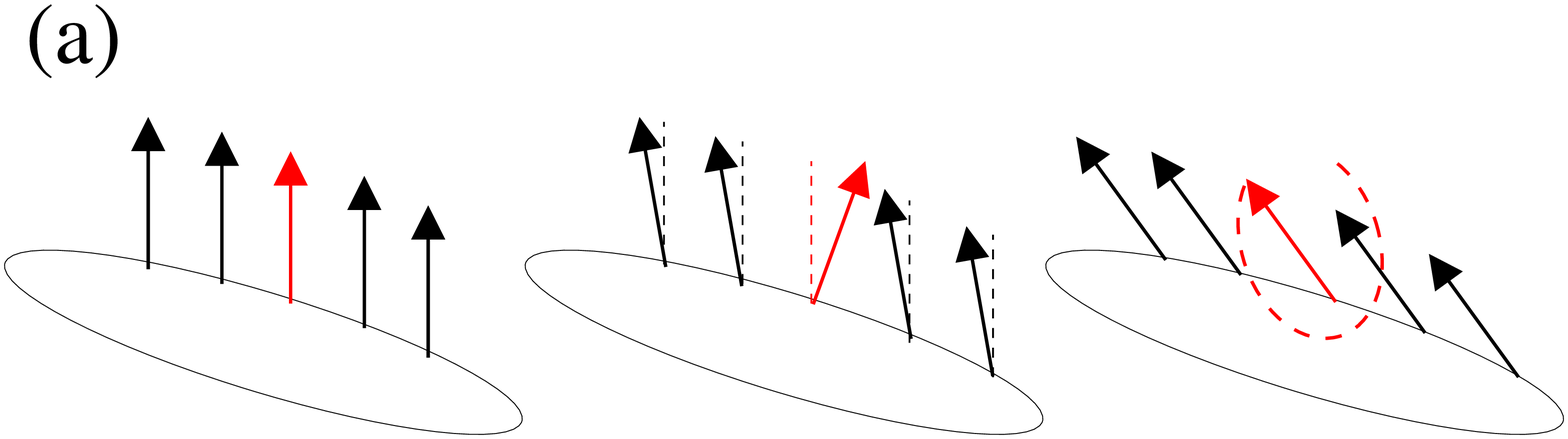}\\[-4mm]
\includegraphics[scale=0.35,angle=270.]{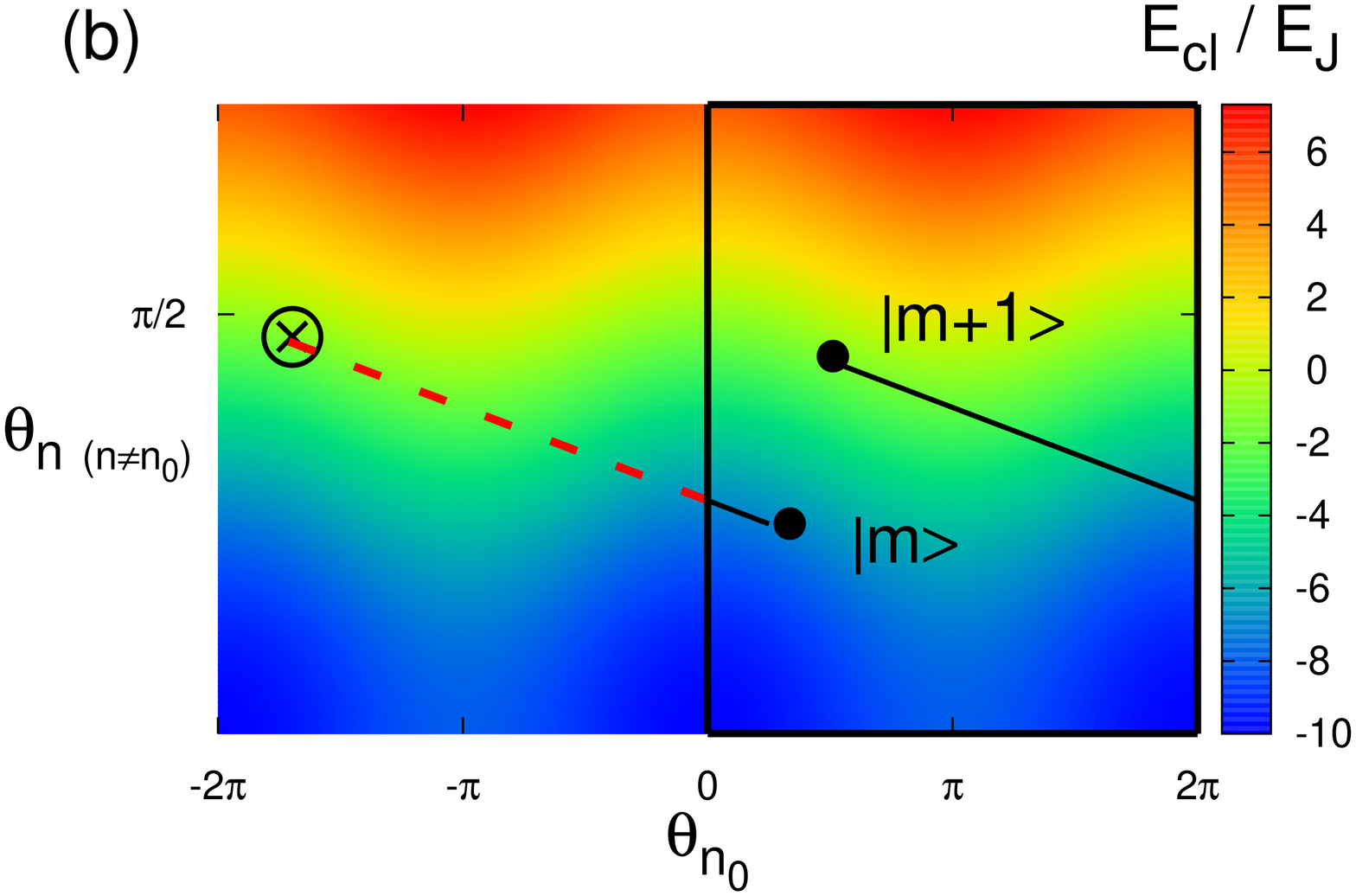}
\caption{Example of a QPS process. {\bf a)} The oval loop represents the JJ ring. The
arrows represent a few phase-differences $\{\theta_n\}$ around the QPS center  at $n_0$,
the phase $\theta_{n_0}$ winding of $2\pi(1-1/N)$ (red arrow). The initial configuration
(left) is the state $\left|m\right>$ and the final configuration (right) is
$\left|m+1\right>$. The central configuration is intermediate between the two states
$\left| m \right>$ and $\left| m+1 \right>$. {\bf b)} The classical energy expressed as
$E_{cl}/E_J=-\cos(x)-(N-1)\cos(y)$ with the
axis $x=\theta_{n_0}-\delta/N$ and $y=\theta_{n} -\delta/N (n \neq n_0)$.
The dots represent the state  $\left| m \right>$ and $\left| m +1 \right>$. The cross
also represents  the state $\left| m +1 \right>$ but in the extended zone
scheme. The black lines represent the physical trajectory of the QPS that connects the
initial and the final state, the (black) solid one in the restricted  zone scheme, the
(red) dashed line in the extended zone scheme. The bold line for the borders marks the
compact region. Parameters: $N=10, \delta=\pi/2, m=1$.}
\label{fig:arrows_PS}
\end{figure}

For finite $C,C_0$, the electrostatic interaction acts as an inertial term on the phases
so that quantum fluctuations occur, giving rise to quantum phase-slips (QPSs).
{\sl At fixed magnetic flux} and in a ring of finite circumference,  the
QPS is a collective process corresponding to the {\sl quantum tunneling in a multidimensional space}
of dimension $N$ between two distinct minima of the potential, corresponding
for instance to the classical states $\left| m \right>$ and $\left| m + 1\right>$, separated by
some energy barrier associated with the potential Eq.~(\ref{eq:E_classical}).
In the multidimensional space, the physical paths $\left\{ \theta_n \right\}$
which connect $\left| m \right>$ and $\left| m \pm 1\right>$ correspond to
a subspace defined by the constraint Eq.~(\ref{eq:constraint}).
Due to this constraint, the multidimensional tunneling reduces to one-dimensional tunneling
in which we have only a few trajectories connecting the initial state and the final state,
see  Fig.~\ref{fig:arrows_PS}.
As it was discussed in Ref.~\onlinecite{MateevLarkinGlazman:2002}, an example of QPS
connecting the states $\left| m \right>$  and $\left| m + 1\right>$ is given by the
displacements
\begin{eqnarray}
\theta_{n_0} &=& \frac{ 2\pi m}{N} \longrightarrow \frac{ 2\pi (m+1)}{N} - 2\pi \, ,
\label{eq:PS_1}\\
\theta_{n} &=& \frac{ 2\pi m}{N} \longrightarrow \frac{ 2\pi (m+1)}{N}  \,\,\,   (n\neq n_0)
\, , \label{eq:PS_2}
\end{eqnarray}
in which the local phase difference $\theta_{n_0}$, $n_0$ being the center of the
QPS, winds by an amount of $2\pi(1+1/N)$ and the whole set of phase differences $\{
\theta_n \}$ $(n \neq n_0)$ shifts in order to preserve the constraint.
Fig.~\ref{fig:arrows_PS}a shows this process.
One can express the classical energy Eq.~(\ref{eq:E_classical})
as $E_{cl}/E_J=-\cos(\theta_{n_0}+\delta/N)-(N-1)\cos(\theta_{n}+\delta/N)$ with
$n \neq n_0$, Fig.~\ref{fig:arrows_PS}b.
%
%
%
%%%%%%%%%%%%%%%%%%%%%%%%%%%%%%%%%%%%%%%%%%%%%%%%%%%%%%%%%%%%%%%%%%%%%%%
%
%
%   FIG.3
%
%

\begin{figure}[hb]
\includegraphics[scale=0.3,angle=270.]{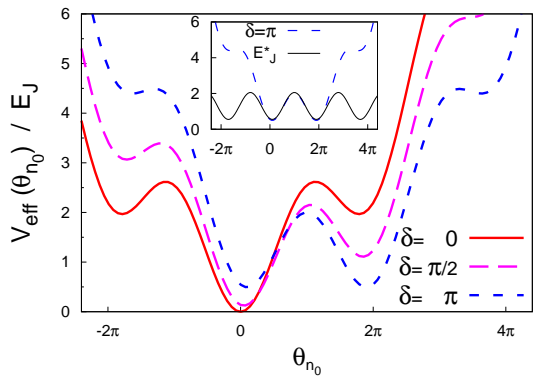}
\caption{
Example of the effective potential $V_{eff}(\theta_{n_0})$
Eq.~(\ref{eq:V_eff}) for $N=10$ and for the values $\delta=0$ solid
(red) line, $\delta=\pi/2$ dashed (purple) line  and $\delta=\pi$ dotted (blue) line.
Inset: comparison between the effective potential at $\delta=\pi$, dotted (blue)
line, with the cosine potential $E_J^* \cos[\theta_{n_0}(N-1)/N]$, solid line (see text).}
\label{fig:Veff}
\end{figure}

As it is shown in Fig.~\ref{fig:arrows_PS}b, the trajectory of the QPS
Eqs.~(\ref{eq:PS_1}),(\ref{eq:PS_2}) can be drawn in a restricted, compact zone
scheme as well as in an extended one.
This fact allows us to introduce the adiabatic potential for the QPS process.
In the limit in which the evolution of the phases is extremely slow,
the kinetic energy can be neglected at any time and one obtains the energy of the system just
minimizing the classical energy for each intermediate configuration  which connects the
initial and the final state.
Hence the adiabatic potential is associated with the line of minimum energy on the surface
Eq.~(\ref{eq:E_classical}) which connects the end points as
shown in the example of Fig.~\ref{fig:arrows_PS}b.
This line is given by  the condition
\begin{equation}
\theta_n= \frac{2\pi m - \theta_{n_0}}{N-1} \,\,\,    (n\neq n_0) \, .
\end{equation}
After a shift of the phase $\theta_{n_0}\rightarrow\theta_{n_0}-\delta/N$,
the effective adiabatic potential $V_{eff}(\theta_{n_0})$ reads
\begin{eqnarray}
\!\!\!\! V_{eff}(\theta_{n_0}) \!\!\! &=& \!\!\!
-E_J \!   \left[ \! \cos(\theta_{n_0})  \!
+ \!
\left(\!N\! -\!1\!\right)
\cos\left(\! \frac{\delta + 2\pi m - \theta_{n_0}}{N-1}  \!\right)
\! \right] \nonumber \\
&\simeq& E_J \!
\left[ - \!
\cos(\theta_{n_0})  \!
+ \!
 \frac{ {\left( \delta + 2\pi m - \theta_{n_0}  \right)}^2 }{2(N-1)} \label{eq:V_eff}
\! \right] \, ,
\end{eqnarray}
where the second, approximate equality is valid for sufficiently long rings $(N \agt 5)$.
At the initial time we have $V_{eff}\left[\theta_{n_0}=(\delta+2\pi m)/N\right]= E_m$ and
at the final time $V_{eff}\left\{\theta_{n_0}=2\pi + [\delta+2\pi (m+1)]/N\right\}=
E_{m+1}$. In Fig.~\ref{fig:Veff}, we show $V_{eff}(\theta_{n_0})$.

In the limit of $N\rightarrow\infty$ we have  $V_{eff}(\theta_{n_0}) = -E_J
\cos(\theta_{n_0})$  and we recover the simplified picture of QPS corresponding to the
quantum tunneling of the phase difference $\theta_{n_0}$ of a given junction $n_0$ from
one minimum of the local Josephson potential $-\cos(\theta_{n_0})$ to the neighboring
one.
%
%
%
%%%%%%%%%%%%%%%%%%%%%%%%%%%%%%%%%%%%%%%%%%%%%%%%%%%%%%%%%%%%%%%%%%%%%%%
%
%
%   FIG.4
%
%

On the other hand, as we explained above, QPS is a collective process
corresponding to a quantum tunneling in a multidimensional space with the constraint
Eq.~(\ref{eq:constraint}).
Consequently, the potential in terms of the variable $\theta_{n_0}$, the center of the QPS,
is {\sl not a periodic} function although the global classical energy of the system
Eq.~(\ref{eq:E_classical}) is $2\pi$ periodically invariant.

\subsection{Effective low-energy description}

We now turn to the effects of quantum phase fluctuations in the limit $E_J \gg E_C, E_0$.
In this case, a simple analysis is possible since the QPS processes occur only rarely,
with an exponentially small amplitude $\nu$, Eq.~(\ref{eq:v_0}).
The single QPS approximation is further analyzed in Sec.~\ref{subsec:discussion} and
Sec.~\ref{subsec:discussion_BKT}, where we estimate its range of validity.

As a consequence of QPSs and for small amplitude $\nu$, the quantum ground state
$\left| \Psi_{GS} \right>$ of the ring corresponds to a superposition of different
classical states of the phases, namely $\left| \Psi_{GS} \right> = \sum_m c_m \left| m \right>$.
The coefficients $c_m$ as well as the quantum ground state energy $E_{GS}(\delta)$
can be obtained from the following effective Schr{\"o}dinger equation\cite{MateevLarkinGlazman:2002}
\begin{equation}
\label{eq:MLG_model}
E_m c_m - \nu_\mathrm{ring} \left( c_{m+1} + c_{m-1} \right) = E_{GS} \, c_m \, ,
\end{equation}
where the term proportional to $\nu_\mathrm{ring} \equiv N \nu$ connects two classical
states differing by a single QPS.
The factor $N$ takes into account the fact a QPS can have the center in
any junction of the chain and this corresponds to different trajectories
in the multidimensional space $N$ so that QPS amplitudes add up coherently.
This coherence has been  recently confirmed experimentally in a short 6-SQUID JJ
chain\cite{Pop:2011b} by the measurements of the Aharonov-Casher interference effect.
The coherence is affected by off-set charge dynamics.
The details of this dynamics are currently not understood.
It is expected to give rise to an additional dependence of $\nu$ on $N$ which is
beyond the scope of the present paper.

\begin{figure}[htbp]
\includegraphics[scale=0.3,angle=270.]{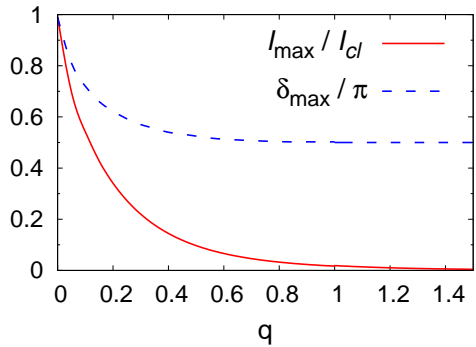}
\caption{Maximum supercurrent (solid line) for the model defined by
Eq.~(\ref{eq:MLG_model}) as a function of $q$, Eq.~(\ref{eq:q-factor}). The dashed line
represents the value of the scaled flux $\delta_\mathrm{max}$ which gives the
maximum supercurrent.}
\label{fig:I_max_q}
\end{figure}

The behavior of the general solution for the ground-state of the model given by
Eq.~(\ref{eq:MLG_model}) is determined by only one dimensionless
parameter
\begin{equation}
\label{eq:q-factor}
q = \frac{N^2 \nu(N)}{2\pi^2 E_J}  \, .
\end{equation}
The ground-state energy $E_{GS}=E_{GS}(\delta;q)$ depends parametrically on $q$, hence
leading to a dependence of the maximum supercurrent $I_\mathrm{max} (q)$ on the
QPS amplitude $\nu$.
This dependence is illustrated in Fig.~\ref{fig:I_max_q} together with the evolution of
the value of the phase $\delta_\mathrm{max}$ that corresponds to the maximum.
For $q \ll 1$, the solution scales as $I_{max}/I_\mathrm{cl} \simeq 1 - (5/2) q^{2/3}$ whereas
it scales as $I_{max}/I_\mathrm{cl}\simeq 24 \sqrt{\pi} q^{4/3} \exp(-8\sqrt{q})$ for $q \gg 1$.

The effective low-energy theory defined by (\ref{eq:MLG_model}) was introduced in
Ref.~\onlinecite{MateevLarkinGlazman:2002} for $C_0=0$ and was discussed for
long JJ rings $N \gg 1$.
As we discussed in sec.~\ref{subsec:SPSap},  the validity of Eq.~(\ref{eq:MLG_model})
extends more generally to include rings of intermediate size and with $C_0 \ne 0$.
In the following, we will obtain the detailed dependence of the QPS amplitude
$\nu_\mathrm{ring}$ on the parameters $N$, $E_J, E_C$, and $E_0$.
Hence, we calculate the ground state energy of the system using
Eq.~(\ref{eq:MLG_model})
to obtain the periodic dependence of $E_{GS}(\delta)$ and $I(\delta)$ on $\delta$,
Eq.~(\ref{eq:I_delta}), from which we extract $I_{max}$.

%
%
%
%%%%%%%%%%%%%%%%%%%%%%%%%%%%%%%%%%%%%%%%%%%%%%%%%%%%%%%%%%%%%%%%%%%%%%%
%
%
%   FIG.5
%
%

\section{QPS in 1D JJ ring}
\label{sec:1PS_approximation}

\subsection{General approach for a single QPS event}
\label{subsec:general_approach}

We present a general approach to calculate the quantum amplitude $\nu$ for
a single QPS event occurring on a ring containing $N$ junctions and with
mutual and ground capacitance $C$ and $C_0$, in the regime $E_J \gg E_C,E_0$.

According to Eqs.~(\ref{eq:PS_1}) and (\ref{eq:PS_2}), the QPS process is a collective
process in which the local phase difference $\theta_{n_0}$ winds by an amount of
$2\pi(1-1/N)$, accompanied by a shift of the whole set of phase differences $\{ \theta_n
\}$ $(n \neq n_0)$.
For rings of circumference $N \agt 5$, the phase differences of the other junctions
will vary only  slightly $\Delta \theta_n \sim 1/N$ compared to the period
$2 \pi$ of the cosine potential.
Hence we can apply the harmonic approximation to describe the
dynamics corresponding of $\theta_n$, $n \neq n_0$.

For vanishing ground capacitance $C_0=0$, the phase-differences of the other $N-1$
junctions behave as independent $LC-$oscillators at the plasma frequency $\omega_p$
whose displacement is inversely proportional to the circumference $N$.
Thus, in the limit of a very large ring $N \gg 1$, the dynamics of the other $N-1$ phase
differences can be neglected\cite{MateevLarkinGlazman:2002} so that the QPS amplitude
$\nu(N)$ can be approximated by the $N-$independent constant value $\nu_0$,
Eq.~(\ref{eq:v_0}).
However, for 1D-JJ rings of finite circumference, the dynamics of the other junctions can have
considerable effects on $\nu$, as we will show below.
Moreover, for finite ground capacitance $C_0 > 0$
the harmonic oscillations of the phase-differences of the $N-1$ junctions play a
crucial role for any ring's circumference.

To calculate the QPS amplitude $\nu$, we start by considering the
partition function associated with the Hamiltonian
Eq.~(\ref{eq:Hamiltonian}), with the constraint Eq.~(\ref{eq:constraint}).
In the path integral formalism, it reads $(\beta=\hbar/k_BT)$
\begin{equation}
\label{eq:Z}
\mathcal{Z}
= \mbox{Tr}\left[ e^{-\frac{\beta}{\hbar}\hat{H}}  \right]
= \prod^{N-1}_{n=0}
\oint \mathcal{D} \left[ \varphi_n(\tau)\right] \,
e^{-  \mathcal{S} / \hbar } \, ,
\end{equation}
where the Euclidean action for the phases $\left\{ \varphi_n(\tau) \right\}$
reads $\mathcal{S} = \int^{\beta}_{0}  d\tau \mathcal{L}$, and $\mathcal{L}$ is the
Lagrangian,
\begin{eqnarray}
\mathcal{L} &=&
\sum^{N-1}_{n=0} \frac{\hbar^2C_0}{8e^2} \dot{\varphi}_n^2
+
\sum^{N-1}_{n=0}  \frac{\hbar^2C}{8e^2}
{\left( \dot{\varphi}_{n+1} - \dot{\varphi}_n \right)}^2
\nonumber \\
& &-\sum^{N-1}_{n=0}
E_J \cos\left( \varphi_{n+1} - \varphi_n  + \frac{\delta_m}{N} \right) \, ,
\label{eqn:Action_phi_n}
\end{eqnarray}
with $\dot{\varphi}_n=d\varphi/d\tau$ and $\delta_m=\delta+2\pi m$.
The compact variables $\{ \varphi_n \}$ are defined on the circle $[0;2\pi[$.
Notice that we shifted the phase differences $\{ \theta_n \}$ with
respect to their average value so that Eq.~(\ref{eq:constraint})
now reads $\sum^{N-1}_{n=0} \theta_n = 0$.
The last constraint is
automatically satisfied by imposing the boundary
condition $\varphi_N=\varphi_0$.

\subsection{Harmonic modes}
\label{subsec:harmonic_modes}

\begin{figure}[h]
\includegraphics[scale=0.3,angle=270.]{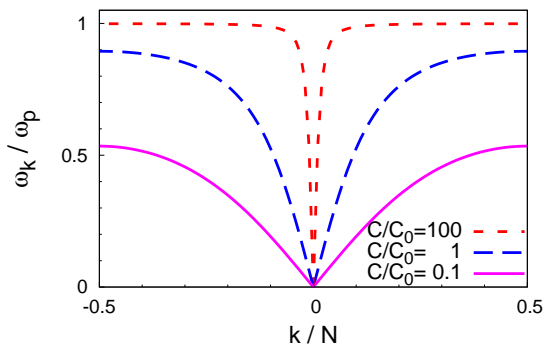}
\caption{Frequency dispersion of the harmonic modes in a JJ ring of $N=100$ junctions.
We use the equivalent index notation $k = -N/2+1,\dots,N/2$.
Solid (purple) line, dashed (blue) line and dotted (red) line are, respectively,
for the ratio $C/C_0=0.1,1,100$. }
\label{fig:frequencies}
\end{figure}

Let us briefly discuss the behavior of the system in the harmonic approximation,
neglecting the QPS.
When the Josephson energy $E_J \gg E_C, E_0$ the phases fluctuate only slightly around
their classical values.
The average phase difference between the neighboring islands is small so that we
can expand the Josephson interaction to lowest (quadratic) order.
The general imaginary-time Lagrangian Eq.~(\ref{eqn:Action_phi_n}) then reduces
to the harmonic one,
\begin{eqnarray}
\mathcal{L}_{har}^{(N)} &=&
\sum^{N-1}_{n=0} \frac{\hbar^2C_0}{8e^2}  \dot{\varphi}^2_n +
\sum^{N-1}_{n=0} \frac{\hbar^2C}{8e^2} {\left( \dot{\varphi}_{n+1} - \dot{\varphi}_{n} \right)}^2 \nonumber \\
& & + \sum^{N-1}_{n=0}
\frac{1}{2}E_J {\left(\varphi_{n+1} - \varphi_{n} \right)}^2
+
\frac{E_J}{2N} \delta^2_m \, ,
\label{eqn:Action_phi_n_harm}
\end{eqnarray}
where we omitted an irrelevant constant term.
Any periodic function $\varphi_n$ defined on the finite
lattice $n=0,\dots,N-1$ can be decomposed as
\begin{equation}
\label{eq:eigenmodes}
\varphi_n =
\frac{1}{\sqrt{N}}
\sum^{N-1}_{k=0} \varphi_k e^{i \frac{2\pi}{N} k n} \, ,
\end{equation}
with the condition for the complex variables
$\varphi_{N-k} = \varphi_{k}^*$ which preserves
the total number of degrees of freedom.
Substituting Eq.~(\ref{eq:eigenmodes}) into
Eq.~(\ref{eqn:Action_phi_n_harm}) and summing over the index $n$, the harmonic
Lagrangian is diagonalized (see also Appendix \ref{app:integration})
\begin{equation}
\label{eq:L_harm_p_k}
\mathcal{L}_{har}^{(N)} =
\sum^{N-1}_{k=0}
\left(
\frac{1}{2} \mu_k
{\left| \dot{\varphi}_k \right|}^2 + \frac{1}{2}  \mu_k \omega_k^2
{\left| \varphi_k \right|}^2
\right)   +
\frac{E_J}{2N} \delta^2_m \, .
\end{equation}
We have introduced the constants
\begin{equation}
\label{eq:mu_k}
\mu_k = \frac{\hbar^2}{4e^2}
\left\{
C_0 + 2 C \left[ 1 -\cos \left( \frac{2\pi}{N} k \right)\right]
\right\} \, ;
\end{equation}
the frequency dispersion is given by
\begin{equation}
\label{eq:omega_k}
\omega_k = \omega_p \; \sqrt{\frac{1-\cos\left(\frac{2\pi k}{N}\right)}
{1-\cos\left(\frac{2\pi k}{N} \right) + \frac{\pi^2}{2\lambda^2} }} \, ,
\end{equation}
where the screening length is $\lambda=\pi\sqrt{C/C_0}$.
For $N\gg 1$, the maximal frequency of the modes $\omega_{max}$ is  given by
\begin{equation}
\label{eq:om_max}
\omega_{max} = 4 \sqrt{ \frac{ E_J}{\hbar^2} \left( \frac{e^2}{4C+C_0} \right) } \, .
\end{equation}
An example of this frequency dispersion is given in Fig.~\ref{fig:frequencies}.
We can distinguish two regimes for $C_0>0$.
For JJ rings longer than the screening length $\lambda$, $N \gg \lambda$,
the spectrum has a linear dispersion for low frequencies.
For shorter rings $N \ll \lambda$, the lowest mode appears almost at the plasma
frequency $\omega_p$ and the linear behavior is completely absent.
For the case $C_0=0$ we recover a flat distribution where all the modes are
degenerate and correspond to the plasma frequency $\omega_p$.
%
%
%
%%%%%%%%%%%%%%%%%%%%%%%%%%%%%%%%%%%%%%%%%%%%%%%%%%%%%%%%%%%%%%%%%%%%%%%
%
%
%   FIG.6
%
%

Let us calculate the phase-dependent ground state energy $E^{(har)}_{GS}(\delta)$ in the
harmonic approximation.
The path integral in Eq.~(\ref{eq:Z}) can be explicitly calculated for the diagonalized
harmonic action Eq.~(\ref{eq:L_harm_p_k}). We obtain
\begin{equation}
\label{eq:Z_har}
\mathcal{Z}_{har}^{(N)} \sim
\left( \prod_{k \neq 0} \frac{1}{2 \sinh\left(\beta \omega_k/2\right)} \right)
e^{-\frac{\beta E_J}{2 \hbar N}
\min_m \delta_m^2 } \, ,
\end{equation}
where the product corresponds to the partition function for an ensemble of $N-1$
independent harmonic oscillators and the exponential factor contains the classical energy
of the system (mod $2\pi$) at finite temperature.
Notice that the zero frequency mode $k=0$ is not involved in the relevant part of the partition
function.
Note also the periodicity of the result Eq.~(\ref{eq:Z_har}) with respect to $\delta$.
The quadratic dependence on $\delta$ in each of the segment $-\pi + 2\pi m < \delta <
-\pi + 2\pi (m+1)$ is a consequence of the harmonic approximation, {\em i.e.},
Eq.~(\ref{eq:E_GS_cl}).

We thus obtain the phase-dependent ground-state energy,
$E^{(har)}_{GS} = -\lim_{\beta=\infty} (\hbar/\beta) \ln \mathcal{Z}_{har}^{(N)}$
\begin{equation}
\label{eq:E_GS_har}
E^{(har)}_{GS} =
\sum_{k \neq 0} \frac{\hbar \omega_k}{2}
+ \frac{E_J}{2N} \min_m \delta_m^2  \, .
\end{equation}
Notice  that to reach the full quantum regime of the harmonic modes, as expressed by Eq.
(\ref{eq:E_GS_har}), the temperature has to be much smaller than the energy of
the lowest mode, $k_BT \ll \hbar\omega_{min}$, which is given by
\begin{equation}
\label{eq:om_min}
\omega_{min} = \omega_{max} \frac{\pi}{N}
\sqrt{
\frac{1+ {\left(  \frac{2\lambda}{\pi} \right)}^2}
{1+ {\left( \frac{2\lambda}{N} \right)}^2 }} \, .
\end{equation}

We conclude that, in the absence of QPS processes, the JJ ring forms a
 closed quantum $LC$-line (each Josephson junction in Fig.~\ref{fig:scheme_array}
is replaced by an inductance $L_J=\hbar^2/4e^2E_J$) formed by $N-1$
independent harmonic oscillators with eigenfrequencies $\omega_k$.
The classical saw-tooth relation between the supercurrent and the total
phase $I=I(\delta)$ is unmodified by the quantum harmonic fluctuations of the phases.
However, harmonic fluctuations are relevant when the QPSs
are restored, as we will now show.

\subsection{Effective QPS action in the presence of harmonic modes}
\label{S_eff_result}

We now turn our attention to the effect of the harmonic modes on the QPS amplitude. We
will restrict our analysis to the limit in which the frequency of the lowest mode
$\omega_{min}$ is greater than the frequency $\nu/\hbar$ associated to the tunneling in
the effective (static) potential $V_{eff}$  Eq.~(\ref{eq:V_eff}). In this adiabatic
limit, the problem can be reduced to an effective two-state problem involving tunneling
between neighboring states $m$ and $m+1$ (see also Sec.~\ref{subsec:discussion_BKT}).

As discussed in Sec.~\ref{subsec:SPSap}, when a single QPS is centered on one specific
junction $n_0$, the dynamics of the other junctions $(n \neq n_0)$ is well described by
the  harmonic approximation.
Then, as seen in Sec.~\ref{subsec:harmonic_modes}, we can consider this part of
the ring as an electromagnetic environment formed by $N-1$ independent harmonic
oscillators and to which the single junction, center of the QPS process, is coupled.
Accordingly, it is natural to cast the full action in a form where the winding
phase-difference $\theta_{n_0}$ is linearly coupled to an ensemble of harmonic
oscillators acting as an external bath.

Without loss of generality, one can assume the center of the QPS to the junction
$n_0=N-1$, namely $\theta = \theta_{N-1} = \varphi_0-\varphi_{N-1}$.
It is useful to introduce the average value of the phase
at the junction $N-1$, namely $\Theta_0=(\varphi_0 + \varphi_{N-1})/2$, so
that one can write
\begin{equation}
\label{eq:new_variables}
\varphi_0    =  \Theta_0 + \frac{\theta}{2} ,
\qquad
\varphi_{N-1}=  \Theta_0 - \frac{\theta}{2}   \, .
\end{equation}
We choose as the relevant variables the set $\mathit{S}$  given by
the winding phase difference  difference $\theta$ together with
the local phases  $\varphi_{n}$ with $n=1,\dots,N-2$.

First we discuss the harmonic expansion of the potential energy
of the Lagrangian Eq.~(\ref{eqn:Action_phi_n}).
In a QPS, the phase-differences across the junctions  remain small with respect to $2\pi$
except at the junction $n_0=N-1$,
\begin{equation}
\sum^{N-1}_{n=0} \cos\left(\! \theta_{n}+\frac{\delta_m}{N}\! \right) \!\simeq
\cos\left(\!\theta + \frac{\delta_m}{N}\!\right) - \frac{1}{2}\sum_{n=0}^{N-2} {\left(\!
\theta_{n} + \frac{\delta_m}{N} \!\right)}^2. \label{eq:exp_cos}
\end{equation}
Using the set $\mathit{S}$, the last sum in Eq.~(\ref{eq:exp_cos}) contains
a quadratic coupling between $\theta$ and the quantities
$\Theta_0-\varphi_{1}$ and $\Theta_0-\varphi_{N-2}$.
Although these terms contain $\theta$ varying by almost $2\pi$, this does not make
invalid our expansion as the overall argument of the cosine function,
representing the phase-difference in the neighboring junctions $n=0$ and $n=N-2$,
remains small during the QPS.

Using the set $\mathit{S}$, the Lagrangian of Eq.~(\ref{eqn:Action_phi_n}) is decomposed as
\begin{equation}
\label{eq:L_single}
\mathcal{L} = \mathcal{L}_{1} + \mathcal{L}_{2}+ \mathcal{L}_{3} \, ,
\end{equation}
where the first term $\mathcal{L}_{1}$ is associated with the winding junction
\begin{equation}
\label{eq:L_1ps}
\mathcal{L}_{1}   \!\!
=
\!\!\frac{\hbar^2( 3C+C_0)}{16e^2} \dot{\theta}^2
- E_J \cos\left( \! \theta + \frac{\delta_m}{N} \! \right)  \!
+  \! E_J
\left(  \! \frac{\theta^2}{4}  \! -  \!\theta\frac{\delta_m}{N}  \! \right) \, .
\end{equation}
The second term $\mathcal{L}_{2}$ describes the environment to which $\theta$ is coupled.
Hereafter we change the notation for the average
\begin{equation}
\Theta_0 \rightarrow \varphi_{0} \, ,
\end{equation}
in order to simplify the following formulas.
Then $\mathcal{L}_{2}$ reads
\begin{eqnarray}
\label{eq:L_bath}
\mathcal{L}_{2} &=& \frac{\hbar^2C_0}{8e^2}  \dot{\varphi}^2_0 +
\sum^{N-2}_{n=0} \frac{\hbar^2}{8e^2}
\left[ C_0 \dot{\varphi}^2_n + C  {\left( \dot{\varphi}_{n+1} - \dot{\varphi}_{n} \right)}^2 \right]
\nonumber \\
&+ & \!\!\! \sum^{N-2}_{n=0}
\frac{1}{2}E_J {\left(\varphi_{n+1} - \varphi_{n} \right)}^2
+ E_J (N-1) {\left( \frac{\delta_m}{N} \right)}^2  ,
\end{eqnarray}
with the periodic boundary condition, $n=0$ corresponding to $n=N-1$.
% where we restored the notation $\Theta_0 \rightarrow \varphi_{0}$
% with the notation $\varphi_{0}=\varphi_{N-1}=\tilde{\varphi}_0$ for the
% periodic boundary condition.
%
Note the extra term in Eq.~(\ref{eq:L_bath}) associated with the (average) phase at $n=0$.
The last term $\mathcal{L}_{3}$ of Eq.~(\ref{eq:L_single})
describes the coupling between the winding junction and the electromagnetic
environment,
\begin{equation}
\label{eq:L_int}
\mathcal{L}_{3} =
\frac{\hbar^2C}{8e^2}
\left( \dot{\varphi}_{N-2} - \dot{\varphi}_{1} \right) \dot{\theta}
+  \frac{E_J}{2} \left( \varphi_{N-2} - \varphi_{1} \right) \theta \, .
\end{equation}
We rewrite the Lagrangians $\mathcal{L}_{2}$ and $\mathcal{L}_{3}$ in terms of the
normal modes $\varphi_k$.
They are given by Eq.~(\ref{eq:eigenmodes}) with $N$ replaced by $N-1$.
In terms of these modes, we have
\begin{equation}
\label{eq:L_bath_diag}
\mathcal{L}_{2} =
\frac{\hbar^2C_0}{8e^2}  \dot{\varphi}^2_{n=0}
+
\mathcal{L}_{har}^{(N-1)}
+
\frac{E_J}{2}
\delta^2_m \left( \frac{N-2}{N(N-1)} \right)
\, ,
\end{equation}
where $\mathcal{L}_{har}^{(N-1)}$ is defined by
Eqs.~(\ref{eq:L_harm_p_k}), (\ref{eq:mu_k}), (\ref{eq:omega_k}) in which we
have to replace $N$ by $N-1$.
Then, from Eq.~(\ref{eq:L_int}), one can see that the
phase $\theta$ is coupled  only to the imaginary part of the modes
$\varphi_k=\varphi_k^R+i\varphi_k^I$  (see Appendix \ref{app:integration})
\begin{equation}
\label{eq:L_int_diag}
\mathcal{L}_{3}= \sum_{k=1}^{k_{max}}  \zeta_k
\left(
\frac{\hbar^2 C}{8e^2} \dot{\varphi}^{I}_k \dot{\theta} +
\frac{E_J}{2}\varphi_k^{I} \theta \right) \, ,
\end{equation}
where $k_{max}=(N-3)/2$ for $N$ odd and
$k_{max}=(N-2)/2$ for $N$ even.
We have introduced the factor $\zeta_k$,
\begin{equation}
\label{eq:zeta_k}
\zeta_k = 4\sin\left[2 \pi k /(N-1)  \right] / {\left(N-1\right)}^{1/2}\, .
\end{equation}
Notice that the mode $k=0$ is not coupled to the winding phase $\theta$ and
we can omit it hereafter.
The partition function associated with the total Lagrangian
Eqs.~(\ref{eq:L_single}), (\ref{eq:L_1ps}), (\ref{eq:L_bath_diag}), (\ref{eq:L_int_diag})
reads
\begin{equation}
\mathcal{Z} \sim
\oint \! \mathcal{D} \! \left[ \theta(\tau)\right] \,
\prod^{k_{max}}_{k=1} \!
\oint \!\! \mathcal{D} \left[ \varphi_k^I(\tau)\right] \,
e^{-\frac{1}{\hbar} \int^{\beta}_0 \!\!\! d \tau
\left( \mathcal{L}_{1} + \mathcal{L}_{2}+ \mathcal{L}_{3}  \right) } \, .
\label{eq:Z_tot_CL}
\end{equation}
It is possible to integrate out the imaginary parts of the $N-2$
harmonic modes to obtain a single effective action describing
the dynamics of $\theta$.
After the calculation, shown in Appendix~\ref{app:integration}, we find
\begin{equation}
\mathcal{Z} \sim \mathcal{Z}^{(N-2)}_{har}
\oint \! \mathcal{D} \!  \left[ \theta(\tau)\right] e^{-S_{eff}[\theta(\tau)] } \, ,
\label{eq:Z_final}
\end{equation}
where $\mathcal{Z}^{(N-2)}_{har}$ is given by
Eq.~(\ref{eq:Z_har}) with $N-1$ replacing $N$.
After a shift of the phase $\theta \rightarrow \theta - \delta_m/N$,
the effective action for the phase $\theta$ is given by
\begin{eqnarray}
& & \!\!\!\!\!\!  S_{eff} \!= \!\!\!\!
\int_0^{\beta} \!\!\!\!\!\! d\tau \!\! \left[\!\!
\frac{\hbar^2}{8e^2}
\!\!
\left( \! \frac{N C}{N-1} \!+ \!\frac{C_0}{2}  \! \right)
\!\!
\dot{\theta}^2
\!\!-\!\!
E_J \! \cos(\theta)
\!+\!\!
\frac{E_J {\left(\delta_m-\theta \right)}^2}{2(N-1)} \!
\right] \nonumber \\
& & \,\,\,\,\, + \frac{1}{2}
\int_0^{\beta} \!\!\!\!\!\! d\tau
\int_0^{\beta} \!\!\!\!\!\! d\tau' \,
G(\tau-\tau') \theta(\tau) \theta(\tau') \, . \label{eqn:S_eff}
\end{eqnarray}
The effective action has a  kernel $G(\tau)$ which is non-local in time
and  whose Fourier series is given by
$G(\tau)=\sum_{\ell} (1/\beta) G_{\ell} \exp(i \omega_{\ell} \tau)$
where $\omega_{\ell}=2\pi\ell/\beta$ are bosonic
Matsubara frequencies and $G_{\ell}$ reads
\begin{equation}
G_{\ell} = \frac{\hbar^2 C_0}{4e^2}
\left[ \frac{\omega_{\ell}^2}{2(N-1)} \right]
\sum^{k_{max}}_{k=1}\frac{1+\cos\left(\! \frac{2\pi k}{N-1} \! \right)}
{1\!\!-\!\!\cos\left(\! \frac{2\pi k}{N-1} \! \right) \!\!+ \!\!
\frac{\pi^2}{2\lambda^2}
\left( \frac{\omega_{\ell}^2}{\omega_{\ell}^2+\omega_{p}^2} \right)
 }
. \label{eq:G_ell}
\end{equation}
The kernel has the relevant property $G_{\ell}=0$ for $\ell = \omega_{\ell} = 0$.
The last relation is equivalent to
$\int^{\beta}_0 \!\! d \tau G(\tau-\tau')=\int^{\beta}_0 \!\!  d \tau' G(\tau-\tau') = 0$.
As a consequence, this term is invariant under a shift of the winding phase
$\theta\rightarrow \theta+ const.$
In other words, upon a proper redefinition  of $G(\omega)=\omega^2 G'(\omega)$,
this (kinetic) term can be written as
$\sim G'(\tau-\tau')\dot{\theta}(\tau) \dot{\theta}(\tau')$.

We observe that  the potential in the first line of $S_{eff}$ in
Eq.~(\ref{eqn:S_eff}) corresponds exactly to the adiabatic potential
$V_{eff}(\theta)$ Eq.~(\ref{eq:V_eff}) introduced in Sec.~\ref{subsec:SPSap}.
This potential breaks formally the periodicity in $\theta$ in the action.
This symmetry breaking is a consequence of the fact that the QPS
is a quantum tunneling in a multidimensional space with the
constraint imposed by the magnetic flux threading the JJ ring
(see discussion in Sec.~\ref{subsec:SPSap}).

In summary, Eqs.~(\ref{eq:Z_final}), (\ref{eqn:S_eff}), and (\ref{eq:G_ell}) constitute
the central result of this paper.
They enable us to calculate the size-dependent QPS amplitude $\nu(N)$ and hence the
phase-dependent ground-state energy and the ring's maximum supercurrent $I_{max}$ in a
broad range of values of the parameters $N, E_J, E_C$ and $E_0$, as we will show in
detail below. However, we first establish a relation with previous
work~\cite{Korshunov:1986,Korshunov:1989} on infinitely long chains by considering the
thermodynamic limit.

\subsection{The thermodynamic limit and the dissipative dynamics}
\label{subsec:S_eff_N=infty}

The effective action Eq.~(\ref{eqn:S_eff}) describes the single winding junction
coupled to its electromagnetic environment constituted by the other $N-2$ junctions
in the harmonic approximation.
This  action is very similar to the one describing the dissipative dynamics of the
single Josephson junction in the framework of the Caldeira-Leggett model.\cite{Caldeira:1981}
In this model, an abstract bath formed by an infinite number of harmonic oscillators is
phenomenologically introduced as the mechanism of irreversible loss of energy in the
Josephson junction.

The external bath  discussed here, expressed by the kernel $G(\tau)$ in Eq.~(\ref{eq:G_ell}),
physically corresponds to the {\sl real harmonic modes} sustained by the Josephson junction ring.
These discrete modes can be experimentally designed and tested.\cite{Haviland:2011}
As long as the ring has finite size, there are a finite number of discrete modes and {\sl
no real dissipation} occurs. We also note that  the interaction between the winding local
phase-difference at the junction $n_0$ and these $N-2$ harmonic modes is characterized by
a linear coupling  through the positions of the oscillators $\varphi_k$ as well as
through their velocities $\dot{\varphi}_k$, see Eq.~(\ref{eq:L_int_diag}). As we will
show now, the difference between the system described by Eqs.~(\ref{eqn:S_eff}),
(\ref{eq:G_ell}) and the standard Caldeira-Leggett model disappears in the limit
$N=\infty$.

Let us consider Eqs.~(\ref{eqn:S_eff}),(\ref{eq:G_ell}).
Taking the limit $N\to \infty$, the first term on the right hand side of
Eq.~(\ref{eqn:S_eff}) reduces to the action for a single capacitively
shunted Josephson junction with a capacitance $C+C_0/2$.
Equation~(\ref{eq:G_ell}) for the kernel $G_{\ell}$ then takes a simple form
by replacing the sum with an integral.
Proceeding in this way, we add the kinetic term of the first line of
Eq.~(\ref{eqn:S_eff}) to $G_{\ell}$  to recover Korshunov's result for the total
kernel $\gamma(\omega_{\ell})$ of a QPS  in a chain of infinite
length,\cite{footnote:detailed-comparison}
\begin{eqnarray}
\gamma(\omega_{\ell}) &=&
\frac{\hbar^2 }{4e^2} \omega_{\ell}^2
\left( C + \frac{C_0}{2} \right) + G_{\ell} \nonumber \\
&=&
E_J{\left(\frac{\omega^2_{\ell}}{\omega_{max}^2} \right)}
\!\!+\!\!
\frac{\hbar \omega_{\ell}}{4}
\sqrt{\frac{E_J}{2E_0}} \sqrt{ 1 + \frac{\omega^2_{\ell}}{\omega_{max}^2}
} \, , \label{eq:G_ell_N=inf}
\end{eqnarray}
where $\omega_{max}$ is defined in Eq.~(\ref{eq:om_max}).
The effective action now reads:
\begin{equation}
S_{eff} =
\int_0^{\beta} \!\!\!\!\!\! d\tau
\int_0^{\beta} \!\!\!\!\!\! d\tau' \,
\frac{1}{2} \gamma(\tau-\tau') \theta(\tau) \theta(\tau')
-
\int_0^{\beta} \!\!\!\!\!\! d\tau
E_J \cos\left[\theta(\tau)\right] \, .
\end{equation}

First, let us discuss the high and low energy regions for the
dynamics of the winding phase difference $\theta$ in the cosine Josephson potential.
These two regions are separated by the condition that the kinetic energy
be respectively larger or smaller than the height of the potential $\sim E_J$.
To estimate the kinetic energy, we determine the effective capacitance of the
junction. This can be achieved by taking the limit $\omega_{\ell} \rightarrow \infty$.
Then the kernel $\gamma(\omega_{\ell})$ Eq.~(\ref{eq:G_ell_N=inf})
corresponds simply to a pure capacitance
\begin{equation}
\gamma(\omega_{\ell}) \simeq \frac{{\left(\hbar \omega_{\ell}\right)}^2}{4e^2}
\left(
C
+
\frac{C_0}{4}
+
\sqrt{\frac{C_0}{4} \left(C + \frac{C_0}{4} \right)}
\right) \, .
\end{equation}
The kinetic energy corresponds to the height of the potential barrier
when $\gamma(\omega_{\ell})=E_J$ giving the threshold $\omega_{\ell} = \omega_{max}$.
For $\omega_{\ell} > \omega_{max}$ the  cosine Josephson potential is a small perturbation
for the dynamics $\theta$.
On the contrary, the range $\omega_{\ell} < \omega_{max}$ corresponds to the energy
region where the winding phase $\theta$ moves well within a potential minimum.
In the absence of the interaction with the modes, its dynamics is harmonic.

The threshold $\omega_{max}$ separates the low- and high-frequency part of the kernel
$\gamma$ Eq.~(\ref{eq:G_ell_N=inf}).
As we have explained, the high-frequency component $\omega_{\ell} > \omega_{max}$ scales
approximately as $\gamma(\omega_{\ell}) \sim \omega^2_{\ell} $ and the chain behaves as a
pure capacitance.
On the other hand, approaching the zero frequency, $\gamma(\omega_{\ell})$ scales linearly.
In a Josephson junction with dissipation, this behavior corresponds to the effect
of an external resistance $R$ leading to a dimensionless damping parameter
$\eta=R_q/R$\cite{Schon:1990} where $R_q=\hbar/4e^2$.
In our model, for $N=\infty$, this resistance corresponds to $R={(L_J/C_0)}^{1/2}$,
the characteristic low-frequency impedance of the chain
related to linear dispersion of the modes.

It is interesting to observe that in the standard Caldeira-Leggett model the two energy
scales --the first one related to the ratio between the height of the potential and the
kinetic  energy and the second one related to the high-energy cut-off for the
dissipation-- are generally different as the latter is determined by the specific
electromagnetic environment to which the junction is coupled.

Defining the relaxation time $\tau_r=R C$, we can
calculate the quality factor given by the ratio between the oscillation period and the
relaxation time,  $Q=2\pi/(\omega_p \tau_r)=2\pi {(C/C_0)}^{1/2}$.
In infinite Josephson junction chains, the underdamped regime
$Q \gg 1$ corresponds thus to $C \gg C_0$.

\section{Results: Vanishing Ground Capacitance}
\label{sec:result_C0=0}

We now go beyond the thermodynamic limit and discuss finite-size effects for a ring in
the limit $C_0=0$. The non-local  kernel then vanishes, $G_{\ell}=0$, and the effective
Lagrangian from Eq.~(\ref{eqn:S_eff}) reduces to
\begin{equation}
\label{eq:L_ef_C}
\mathcal{L}_{eff} =
\left(\frac{N}{N-1}\right) \frac{\hbar^2C}{8e^2}  \dot{\theta}^2 -
E_J \cos(\theta) +  \frac{E_J {\left(\delta_m-\theta \right)}^2}{2(N-1)}.
\end{equation}
This result has a simple interpretation.
Let us write the total Lagrangian, Eq.~(\ref{eq:L_single}), with $C_0=0$ in
terms of the phase differences $\theta_n$,
\begin{equation}
\label{eqn:Action_theta_n_2}
\mathcal{L} =
\frac{\hbar^2C}{8e^2}  \dot{\theta} -  E_J \cos(\theta)
+
\sum^{N-1}_{n=1}
\left(
\frac{\hbar^2C}{8e^2}  \dot{\theta}_n
+ \frac{E_J}{2} \theta^2_n
\right) \, .
\end{equation}
We observe that the phases $\{ \theta_n \}$ are not coupled explicitly to the tunneling
phase $\theta$. But the total phase difference is fixed $\sum_n \theta_n + \theta =
\delta_m$ which enforces an implicit interaction between them. In case of identical
phase-differences $\theta_n$ for all $N-1$ junctions, we have
\begin{equation}
\label{eqn:theta_n_d}
\theta_n = \frac{\delta_m-\theta}{N-1}  , \qquad
\dot{\theta}_n = -\frac{\dot{\theta}}{N-1}  \, .
\end{equation}
By inserting Eq.~(\ref{eqn:theta_n_d})
 in the  action Eq.~(\ref{eqn:Action_theta_n_2}),
we obtain the effective action Eq.~(\ref{eq:L_ef_C}).
Clearly, in the limit $N \rightarrow \infty$ finite size corrections vanish and we
recover the simple action for the single JJ.

A finite value for $N$ first of all modifies the kinetic term.
The constraint-induced
coupling to the other junctions increases the inertial mass $C^*/C$ of the single phase
performing an almost complete winding, leading to a reduction of the charging energy,
\begin{equation}
\label{eq:E_C_ren}
\frac{C^{*}}{C} =  \frac{N}{N-1} \, , \qquad \frac{E^{*}_C}{E_C} = 1-\frac{1}{N}  \, .
\end{equation}
This causes a reduction of the QPS amplitude.

On the other hand, as we can see from Eq.~(\ref{eq:L_ef_C}), the action of the phase
$\theta$ involves the effective adiabatic potential $V_{eff}(\theta)/E_J = 1 - \cos\theta +
{(\delta-\theta)}^2/(N-1)/2$, see Fig.~(\ref{fig:Veff}), which is not purely sinusoidal
and depends on $N$, leading to additional finite-size corrections.

Before discussing these corrections, we note that for short rings,
the effective potential can have only one minimum for arbitrary value of $\delta$.
This is the case, for instance, when $\delta=0$ for $N<5$.
In this regime the harmonic approximation fails too.
Focusing on lengths longer than $N \agt 5$, the potential has always two
minima for which the QPS process is well defined.

Due to the periodicity in $\delta_m$ and the symmetry in $(\delta_m \leftrightarrow
-\delta_m)$ for the quantum ground-state of the system, we can restrict our discussion
to values of $\delta$ between $0$ and $\pi$ $(m=0)$.
In this case, the relevant tunneling process corresponds to an increase of
the variable $\theta$.
The two positions corresponding to the two minima of the effective potential  are,
respectively, $\theta_l=\delta/N$ and $\theta_r=2\pi+(\delta-2\pi)/N$, with the energies
$V_{eff}(\theta_l)$ and $V_{eff}(\theta_r)$ which correspond to the two classical
energies associated with the initial state and the final state before and after a
QPS, respectively.

\subsection{Symmetric effective potential $(q<1)$}

For small values of the parameter $q$, Eq.~(\ref{eq:q-factor}), the hopping term is
small, Eq.~(\ref{eq:MLG_model}).
The QPS process couples mainly two neighboring classical states, $\left| m \right>$ and $\left| m+1 \right>$,
and it is significant when they are degenerate.
This occurs for $\delta \simeq \pi$ which corresponds to the point of
the maximum of the supercurrent in this regime (Fig.~\ref{fig:I_max_q}).

Therefore, we start by discussing the case $\delta=\pi$ when
the effective potential $V_{eff}(\theta)$ is symmetric around the maximum,
Fig.~(\ref{fig:Veff}).
Then the effective potential is very well approximated
by a renormalized cosine potential of the form
$- E_J^* \cos[\theta (N-1)/N]$ (see inset Fig.~\ref{fig:Veff}) where $E_J^*$ is
the renormalized Josephson energy (half of the height of the energy barrier separating the two wells),
\begin{equation}
\label{eq:E_J_ren}
\frac{E_J^*}{E_J} = \frac{1}{2}
\left[
1+\cos\left( \frac{\pi}{N}\right)
- \frac{\pi^2}{2N^2}
\left(N-1\right)
\right]
\simeq
1 - \frac{\pi^2}{4N} \, .
\end{equation}
The ratio $E_J^*/E_J$  is an increasing  function of $N$ which converges to one
in the limit of $N \gg 1$.
Thus for any finite-size system, this correction corresponds to
a decrease of the barrier for quantum tunneling which therefore would lead to
an enhancement of the QPS amplitude, in contrast to the effect
of the renormalized capacitance, Eq.~(\ref{eq:E_C_ren}).
Another effect which enhances the tunneling amplitude is the reduction of the distance
between the two minima of the effective potential,
\begin{equation}
\label{eq:theta_ren}
\frac{\Delta \theta^*}{2\pi} = 1 - \frac{1}{N} \, ,
\end{equation}
contributing to the enhancement of the tunneling amplitude for finite $N$.
Similar renormalization effects have been discussed in Ref.~\onlinecite{Catelani:2011},
albeit for different superconducting circuits.

\begin{figure}[hbtp]
\includegraphics[scale=0.3,angle=270.]{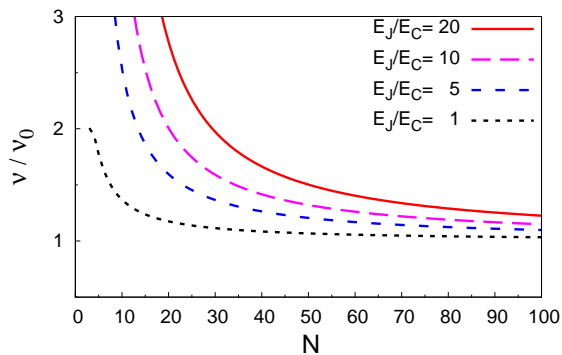}
\caption{For $C_0=0$, the renormalized QPS amplitude
 $\nu$ as a function of $N$, Eq.~(\ref{eq:v_ren}), scaled with $\nu_0$,
Eq.~(\ref{eq:v_0}), for different values of the ratio $E_J/E_C$.}
\label{fig:v_ren}
\end{figure}

The three different effects given by
Eqs.~(\ref{eq:E_C_ren}), (\ref{eq:E_J_ren}), (\ref{eq:theta_ren}) combine in the final
expression for the renormalized effective amplitude for the quantum tunneling between the
two degenerate minima,
\begin{equation}
\label{eq:v_ren}
\nu(N) = \sqrt{\left(\frac{2\pi}{\Delta\theta^*}\right) }
\frac{4}{\sqrt{\pi}}
{\left( 8 E^{*3}_J E^*_C \right)}^{\frac{1}{4}}
e^{- \left( \frac{\Delta\theta^*}{2\pi} \right)  \sqrt{8  \frac{E^{*}_J}{E^*_C}}} \, .
\end{equation}
For the range of interest $N\geq 5$, the QPS amplitude Eq.~(\ref{eq:v_ren})
decreases with the length $N$  indicating that the latter effects
(reduced barrier height and distance between minima) dominate the capacitance
renormalization.
Indeed, when $\nu$ is scaled with the amplitude $\nu_0$ (corresponding to the limit
$N=\infty$), the leading exponential term reads  $\nu/\nu_0 \sim \exp(N_c/N)$ with $N_c
=(\pi^2 + 4) \sqrt{8E_J/E_C}$. Here $N_c$ is the typical length below which this
finite size correction becomes relevant.
The behavior of the ratio $\nu/\nu_0$ is shown in Fig. \ref{fig:v_ren}.
%
%
%
%%%%%%%%%%%%%%%%%%%%%%%%%%%%%%%%%%%%%%%%%%%%%%%%%%%%%%%%%%%%%%%%%%%%%%%
%
%
%   FIG.7
%
%

\begin{figure}[htbp]
\includegraphics[scale=0.3,angle=270.]{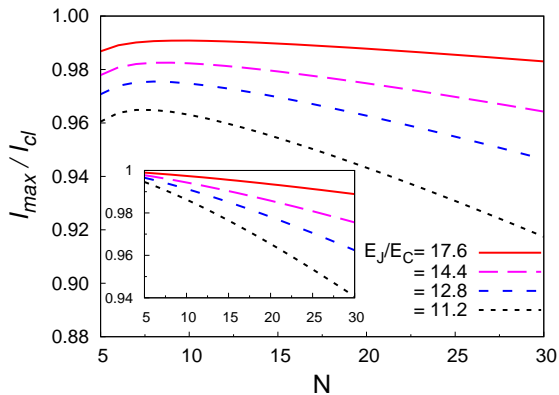}
\caption{The maximum supercurrent $I_{max}$ scaled with $I_\mathrm{cl}$
as a function of $N$ and
at different values of the ratio $E_J/E_C$ by using the QPS amplitude $\nu(N)$
Eq.~(\ref{eq:v_ren}). Inset:  the maximum supercurrent by using the QPS amplitude
$\nu_0$, Eq.~(\ref{eq:v_0}).}
\label{fig:I_ren_1}
\end{figure}

The effective QPS amplitude $\nu$ decreases with increasing $N$. As a result the
parameter $q \sim N^2 \nu(N)$ defined in Eq.~(\ref{eq:q-factor}) has a non-monotonic
behavior as a function of $N$.
Accordingly, the maximum supercurrent obtained for
Eq.~(\ref{eq:MLG_model}) has also a non-monotonic dependence on
the total length $N$.
In Fig.~\ref{fig:I_ren_1} we show the maximum supercurrent as a function of the
ring size $N$ for different ratios of $E_J/E_C$.
Decreasing $E_J/E_C$, the non-monotonic behavior occurs at shorter lengths.
%
%
%
%%%%%%%%%%%%%%%%%%%%%%%%%%%%%%%%%%%%%%%%%%%%%%%%%%%%%%%%%%%%%%%%%%%%%%%
%
%
%   FIG.8
%
%

\subsection{Asymmetric effective potential $(q \agt 1)$}

When the parameter $q$ is of order $\sim 1$, the maximum supercurrent for the
model given by Eq.~(\ref{eq:MLG_model}) shifts from the phase difference
$\delta = \pi$ to the phase difference $\delta = \pi/2$, Fig.~\ref{fig:I_max_q}.
The current-phase relation is strongly modified passing from a sawtooth to a sinusoidal
function.\cite{MateevLarkinGlazman:2002} We should then solve Eq.~(\ref{eq:MLG_model})
using the renormalized QPS amplitude $\nu_{as}$ associated with the full asymmetric
effective potential $V_{eff}$ and which depends on $\delta$, Fig.~\ref{fig:Veff}.

To reach the regime $q\agt1$ there are two possible ways.
First, it can be reached by decreasing  the ratio $E_J/E_C$ at given fixed length $N$,
which is outside the range of validity of the present work. Alternatively, for a given
ratio $E_J/E_C$, we can increase the length $N$ of the system. The finite size effects
that we have discussed in the previous section for the phase difference $\delta=\pi$
vanish as $1/N$.
We now show  that the finite-size corrections due to the asymmetry of the barrier vanish
more rapidly, namely as $1/N^2$. Consequently, in the cross-over range in which
$N$ spans from $N_{min}=6$ to $N \gg 1$, we can neglect the difference between $\nu_{as}$
(for $\delta \neq \pi$) and $\nu$  (for $\delta =\pi$). It is then justified to use the
renormalized amplitude $\nu$ for the symmetric potential, Eq.~(\ref{eq:v_ren}), when
calculating the maximum supercurrent for the model Eq.~(\ref{eq:MLG_model}) in the full
range of $q$.

Let us introduce $\kappa= 1-\delta/\pi$ as the natural parameter to quantify the
asymmetry.
For an asymmetric potential, we can approximate the hopping energy between the two levels
as the geometrical average obtained by considering the hopping for the left part of the
potential and the right part of the potential with respect to
the maximum,\cite{Rastelli:2012,asymmetric-double-well}
\begin{equation}
\label{eq:v_asym}
\nu \simeq \sqrt{\nu_L \nu_R} \, .
\end{equation}

The left (right) amplitude $\nu_{L}$ $(\nu_R)$ is the tunneling amplitude for the
symmetric double potential,  Eq.~(\ref{eq:v_ren}), with the
barrier height $2 E_{J}^{(+)}$ ($2 E_{J}^{(-)}$) given by the difference between the maximum energy and the
left (right) minimum. In  a similar way, $\Delta\theta^{(\pm)}/2\pi$ is the distance between
the maximum and the left (right) minimum point.  Keeping the leading term of the
expansion in $1/N$, they read $E_{J}^{(\pm)}/E_J = 1 - (\pi^2/4N) \pm (\pi^2/2N) \kappa $ and
$\Delta\theta^{(\pm)}/2\pi = 1 - 1/N \pm (\pi/N )\kappa$ where $\pm$ is for $L$ and $R$.
Inserting this expansion into Eq.~(\ref{eq:v_asym}), we finally conclude that
 finite-size corrections associated with the asymmetry parameter $\kappa$ cancel at order $1/N$.
When higher-order corrections are taken into account by using the full formula
Eq.~(\ref{eq:v_asym}) for the QPS amplitude $\nu=\nu(\delta)$, the differences
with the amplitude at $\nu=\nu(\pi)$ are practically unnoticeable.

Finally, our discussion is valid in the limit in which the asymmetry is sufficiently weak
such that the two level description remains valid.
Close to the resonant condition, the excited and ground harmonic levels
between two neighboring states in the effective potential, Fig.~\ref{fig:Veff},
are almost degenerate.
This implies the inequality $E_{n}-E_{n+1} < \hbar \omega_p$ which translates into
an interval for the magnetic flux  $|\Delta \delta| < N\sqrt{2E_C/E_J}/\pi$
around the degeneracy point at half a flux quantum.
As long as the maximum supercurrent is well within this interval,
our analysis based on the simple two level description holds.
Note that in the limit $N \rightarrow \infty$, the range $|\Delta \delta|$ is quite large
as a consequence of the fact that  $V_{eff}$  in Fig.~\ref{fig:Veff} is
practically indistinguishable from the periodic and multi-degenerate cosine potential.
On the other hand,  for short rings, the maximum supercurrent occurs around
$\delta \simeq \pi$, Fig.~\ref{fig:I_max_q}, which is well inside the interval $|\Delta \delta|$
for a suitable choice of the parameters.

\section{Effects of the capacitance $C_0$}
\label{sec:results_general}

We now consider the effect of the ground capacitance $C_0$ on the QPS amplitude. In view
of the discussion of the previous section, we will focus on the analysis of the case
$\delta=\pi$.
Then the effective potential appearing in the first line of Eq.~(\ref{eqn:S_eff}) is
replaced by a renormalized cosine potential
\begin{eqnarray}
\label{eqn:S_eff_C0}
S_{eff} &=& \int_{0}^{\beta} \!\!\!\!\!\! d\tau
\left[ \frac{\hbar^2}{8e^2} \left(\! C^*\!\!+\!\!\frac{C_0}{2}\! \right)
\dot{\theta}^2(\tau)
\!\!-
\!\!
E_J^* \cos\left( \frac{2\pi}{\Delta \theta^*} \,  \theta(\tau)  \right) \right] \nonumber \\
&+& \frac{1}{2}
\int_{0}^{\beta} \!\!\!\!\!\! d\tau
\int_{0}^{\beta} \!\!\!\!\!\! d\tau'
G(\tau-\tau') \theta(\tau)\theta(\tau') \, ,
\end{eqnarray}
where $C^*$, $E_J^*$ are defined in Eqs.~(\ref{eq:E_C_ren},\ref{eq:E_J_ren}) and
$\Delta \theta^*$ in Eq.~(\ref{eq:theta_ren}).
Within the semiclassical instanton approach,\cite{Kleinert:1995}
the QPS amplitude reads
\begin{equation}
\label{eq:nu_general} \nu = A \exp\left[ - S^{(cl)}_{eff}/\hbar \right],
\end{equation}
where $S^{(cl)}_{eff}$  is the effective action  Eq.~(\ref{eqn:S_eff_C0})  evaluated at
$\theta_{cl}(\tau)$, the asymptotic path which minimizes the  action and which connects
the two relevant minima in the limit $\beta\rightarrow\infty$, {\em i.e.}, the instanton
solution. The prefactor $A$ is related to the quantum fluctuations around this minimum
path.\cite{Kleinert:1995}

In contrast with the previous analysis for $C_0=0$, we now have to take into account the
effects of the kernel $G(\tau)$ in the action Eq.~(\ref{eqn:S_eff_C0}).

\subsection{Parabolic approximation}
\label{sec:par}
The first step is to find the classical path $\theta_{cl}(\tau)$.
For a cosine-potential, this solution is known analytically only when the kernel is zero
$G(\tau)=0$ ($C_0=0$).
For the general case, to the best of our knowledge, the solution is unknown.
We use the Villain approximation to solve the problem.\cite{Villain:1975}
We replace the periodic cosine potential by a parabolic potential
\begin{eqnarray}
\label{eq:S_eff_parabol}
S_{p} &=& \int_{-\frac{\beta}{2}}^{\frac{\beta}{2}} \!\!\!\!\!\! d\tau
\left[  \frac{\hbar^2}{8e^2}
\left(\! C^*\!\!+\!\!\frac{C_0}{2}\! \right) \!\!
\dot{\theta}^2 \!\!  +  \!\!
\frac{V_J}{2} \min_m
{\left( \theta - m \, \Delta \theta^* \right)}^2
\right] \nonumber \\
&+& \frac{1}{2}
\int_{-\frac{\beta}{2}}^{\frac{\beta}{2}} \!\!\!\!\!\! d\tau
\int_{-\frac{\beta}{2}}^{\frac{\beta}{2}} \!\!\!\!\!\! d\tau'
G(\tau-\tau') \theta(\tau) \theta(\tau') \, .
\end{eqnarray}
It is worth noting that, for $G=0$ $(C_0=0)$ and $N \gg 1$, the instanton solution for
this potential yields the action $S^{(cl)}_{eff}/\hbar = \pi^2 (V_J/8E_C)^{1/2}$, {\em
i.e.}, the numerical coefficient is different from the one found for a cosine potential
with the same amplitude $V_J$.
Thus, in order to recover the previous results for
$C_0=0$, it is convenient to set the height of the parabolic periodic potential  to the
value $V_J = (8/\pi^2)^2 E_J^*$ to take into account the difference between the profiles
of the two potentials.

\begin{figure}[hbtp]
\includegraphics[scale=0.3,angle=270.]{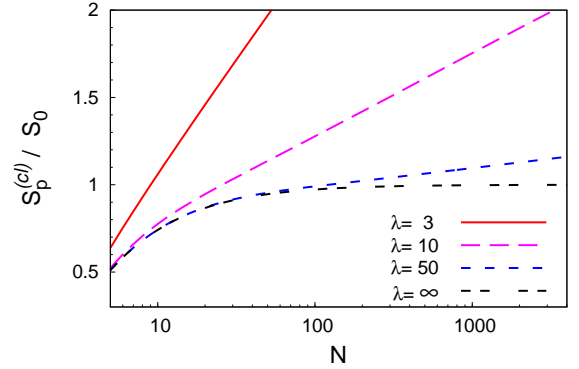}
\caption{For $E_J=8E_C$, the classical action $S_{p}^{(cl)}$
Eq.~(\ref{eq:S_par}) scaled with $S_0 = (8 E_J/E_C)^{1/2}$ as
a function of $N$ for different screening lengths $\lambda = 3,10,50,\infty$.
Notice the cross-over at $N \agt \lambda$ to the logarithmic scaling $\sim \log(N)$.}
\label{fig:S_par}
\end{figure}

After the calculation (see Appendix \ref{app:parabolic} for details), the action
Eq.~(\ref{eq:S_eff_parabol}) with the instanton path
reads~\cite{note:parabolic_approximation}
\begin{equation}
\label{eq:S_par}
S^{(cl)}_p =
2\pi V_J
{\left(\!1 - \frac{1}{N} \!\right)}^2
\int^{\infty}_{0}  \!\!\!\!\!\!\! d\omega
\frac{1}{\omega^2+
\frac{4e^2V_J/\hbar^2}{C^*+\frac{1}{2}C_0+\frac{4e^2}{{(\hbar\omega)}^2}G(\omega)} }\, ,
\end{equation}
where $G(\omega)$ is the continuous limit for the Fourier transform
$G(\omega_{\ell})=G_{\ell}$ defined in Eq.~(\ref{eq:G_ell}).
In Fig.~\ref{fig:S_par}, the behavior of $S^{(cl)}_p$ is shown as a function
of the ring size $N$.
For $C_0=0$ and $N \gg 1$, $S^{(cl)}_p$ saturates to $S_0= (8E_J/E_C)^{1/2}$.
When the ground capacitance is restored, we find a logarithmic scaling
with $N$ for $N > \lambda$.
Specifically, we find that $\nu \sim \exp[-\alpha\log(N)] =
1/N^{\alpha}$ with $\alpha=\pi\sqrt{E_J/(8E_0)}$
(see Appendix \ref{app:analytic}).
%
%
%
%%%%%%%%%%%%%%%%%%%%%%%%%%%%%%%%%%%%%%%%%%%%%%%%%%%%%%%%%%%%%%%%%%%%%%%
%
%
%   FIG.9
%
%

\subsection{The prefactor A}

So far our semiclassical approach was general.
We will now restrict ourselves to the calculation of the finite-size corrections entering
the renormalized amplitude $\nu$ with exponential accuracy; this corresponds to the
leading dependence on $N$.
Correspondingly, we will use an approximate expression for the prefactor $A$.
This prefactor is associated with the quantum Gaussian fluctuations around the classical
path.
Specifically, we neglect the contribution of the low energy paths
having a mean kinetic energy lower than the height of the potential.

Proceeding as in Sec.~\ref{subsec:S_eff_N=infty},  to estimate
the kinetic energy we determine the effective capacitance of the junction in the ring
but now taking into account the finite size effects.
For $\omega_{\ell} \rightarrow \infty$ the kernel Eq.~(\ref{eq:G_ell})
reduces to $G_{\ell} =\Delta C {(\hbar\omega_{\ell})}^2/(4e^2)$ where
\begin{equation}
\label{eq:Delta_C}
\Delta C
=
\frac{C_0}{2\left(N-1\right)}
\sum_{k=1}^{k_{max}} \frac{1+\cos\left(\! \frac{2\pi k}{N-1} \! \right)}
{1\!\!-\!\!\cos\left(\! \frac{2\pi k}{N-1} \! \right) \!\!+ \!\!
\frac{\pi^2}{2\lambda^2} } \, ,
\end{equation}
so that the effective capacitance of the junction corresponds to
\begin{equation}
\label{eq:C_eff}
C_{eff} = C^* + \frac{1}{2}C_0 + \Delta C \, .
\end{equation}
Again, from Eq.~(\ref{eq:S_eff_parabol}), as $V_J \sim E_J$
we see that the threshold frequency separating the high and low energy
regions is still $\omega_{\ell} =\omega_{max}$ for $C > C_0$.
From these observations, in order to take into account the contribution of the high
energy quantum fluctuations, we replace the effective capacitance in the prefactor $A$ of
Eq.~(\ref{eq:v_ren}) (case $C_0=0$), with the effective capacitance Eq.~(\ref{eq:C_eff}).
The result reads
\begin{equation}
\label{eq:A_factor}
A = \frac{4}{\sqrt{\pi}} \sqrt{\left(\frac{2\pi}{\Delta\theta^*}\right) }
 {\left(8 E^{*,3}_J \frac{e^2}{2C_{eff}} \right)}^{\frac{1}{4}}.
\end{equation}

We now discuss the conditions for this
approximation to be valid. For $\omega_{\ell} < \omega_{max}$ the kernel $G$ in
Eq.~(\ref{eq:S_eff_parabol}) couples the dynamics of the winding phase to the one of the
modes.
As we have seen at the end of Sec.~\ref{subsec:S_eff_N=infty} at
$N=\infty$, for $C \gg C_0$ this interaction between the winding phase and the modes
corresponds to a perturbation at low energies for the Gaussian  harmonic
fluctuations.
Indeed, at $N=\infty$ we have weak damping for the quality factor
$Q = 2\pi {(C/C_0)}^{1/2} \gg 1$.
Although at finite $N$ the dynamics of the JJ ring does not correspond to a real
resistance, the ratio $C_0/C$ still plays  the role of a dimensionless coupling  between
the  winding  phase and the harmonic bath.
The effect of this interaction on the low energy Gaussian fluctuations can be
neglected in the prefactor but not in the exponent  where, as we have seen, they strongly
affect the instanton classical path.

\subsection{Superconductor-Insulator transition}

By inserting Eq.~(\ref{eq:S_par}) and Eq.~(\ref{eq:A_factor}) into
Eq.~(\ref{eq:nu_general}), we obtain the QPS amplitude which we use to calculate
the  maximum supercurrent.
Here we discuss some numerical results valid in a general
range of parameters while in Appendix \ref{app:analytic} we discuss some analytic
results valid for very long chains $(N \gg \lambda)$.

\begin{figure}[hbtp]
\includegraphics[scale=0.34,angle=270.]{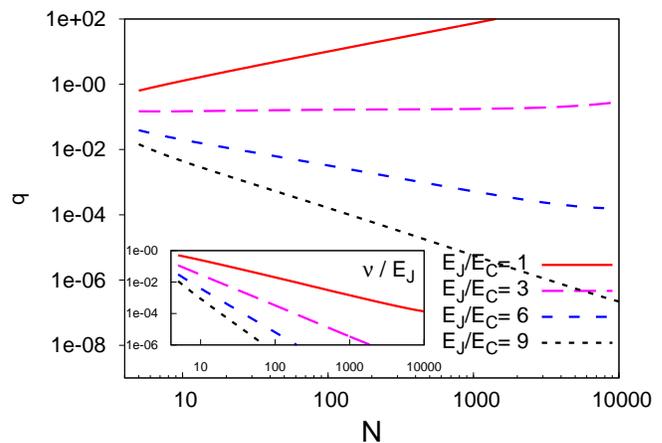}
\caption{The parameter $q$ as a function of $N$ for $C=2 C_0$ and different
ratios $E_J/E_C$. Inset: the behavior of the QPS amplitude $\nu$ scaled with
$E_J$.}
\label{fig:q_ren}
\end{figure}

In Fig.~\ref{fig:q_ren}, we plot the factor $q=N^2\nu(N)/(2\pi^2)$ as a function of the
ring size $N$ for the values $C=2 C_0$ and for different ratios of
$E_J/E_C$. In the inset of Fig.~\ref{fig:q_ren} we also show the behavior of $\nu/E_J$.
The first striking observation is that for large ratios of $E_J/E_C = 2 (E_J/E_0)$,
{\sl the parameter $q$ does not increase as $N^2$} but, indeed, it vanishes upon
increasing the length.
This is due to the logarithmic dependence of the action $S_{eff}$ that we have obtained
in Sec.~\ref{sec:par}.

The presence of the logarithmic dependence is related to the lowest energy modes (see
Appendix \ref{app:analytic}). Indeed, below the corresponding threshold
$\omega_{\ell} \ll \omega_{min}$, the winding junction feels the discreteness of the
spectrum of the environment with which it can exchange energy. It is equivalent
to say that there is no real dissipation at low frequency.
An instanton solution conserving the initial energy still exists with a finite
value of the corresponding action.
For $N \gg \lambda$, we have $\omega_{min} \sim {(8E_JE_0)}^{1/2}\pi/N$.
Eventually,  $\omega_{min}$  vanishes as $N\rightarrow\infty$
and the action associated  to the instanton diverges.

As a consequence, the parameter $q$ behaves as $q \sim N^2 \nu \sim N^{2-\alpha}$ and it
scales either to zero  or to infinity for $N=\infty$. A cross-over is therefore expected
at  some critical  ratio $E_J/E_0$ as $\alpha=\pi\sqrt{E_J/(8E_0)}$
(see Appendix \ref{app:analytic}).
%
%
%
%%%%%%%%%%%%%%%%%%%%%%%%%%%%%%%%%%%%%%%%%%%%%%%%%%%%%%%%%%%%%%%%%%%%%%%
%
%
%   FIG.10
%
%

\begin{figure}[htbp]
\includegraphics[scale=0.34,angle=270.]{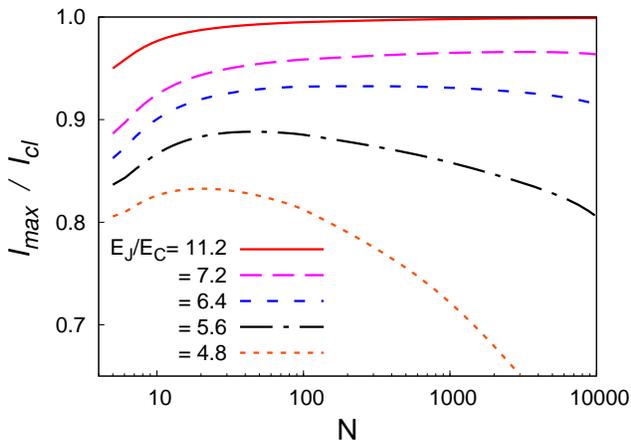}
\caption{The maximum supercurrent $I_{max}$ scaled with the classical
value $I_\mathrm{cl}$ as a
function of $N$ for $C=2 C_0$ and different ratios $E_J/E_C$.}
\label{fig:I_ren}
\end{figure}

The behavior of the maximum supercurrent $I_{max}$ scaled with $I_\mathrm{cl}$ as a function of
$N$, Fig.~\ref{fig:I_ren}, is related to the behavior of  $q$.
{\sl In the very long length limit},
$I_{max}/I_\mathrm{cl}$ increases and saturates to one for $E_J/E_0 > 3.24$
whereas for $E_J/E_0 < 3.24$ it vanishes. On the other hand, we remark that for finite
systems, the current shows a non-monotonic behavior: it increases with $N$ up to a
maximum after which it decreases.\vspace{-1mm}
%
%
%
%%%%%%%%%%%%%%%%%%%%%%%%%%%%%%%%%%%%%%%%%%%%%%%%%%%%%%%%%%%%%%%%%%%%%%%
%
%
%   FIG.11
%
%

The interplay between finite size effects, which  scale  as  $\nu(N) \sim
\exp(N_c/N)$, and the low energy modes at finite $C_0$,
which reduce the QPS amplitude as $\nu \sim 1/N^{\alpha}$, cause that the total
QPS amplitude of the ring $\nu_{ring} = N \nu(N)$ has a weak $N$-dependence  so
that  the dependence of the maximum supercurrent on $N$ appears almost flat over a large
range of $N$ when the critical ratio  is approached (see Fig. \ref{fig:I_ren}).

\begin{figure}[hbtp]
\includegraphics[scale=0.32,angle=270.]{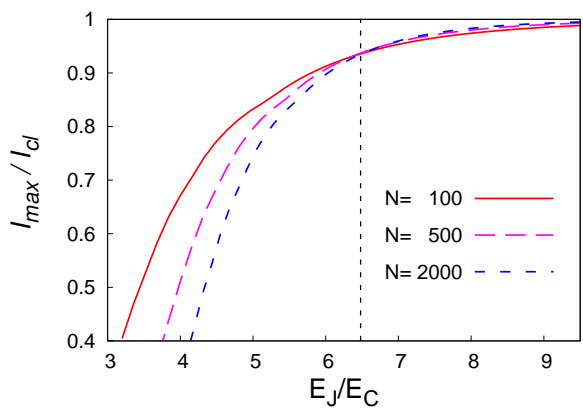}
\caption{$I_{max}$ scaled with $I_\mathrm{cl}$ as a function of $E_J/E_C =2 E_J/E_0$
for rings of different circumferences. The dashed vertical line
corresponds to the critical value  $E_J/E_0=32/\pi^2$  of
Ref.~\onlinecite{Korshunov:1989} for $C = 2 C_0$.}
\label{fig:cross-over}
\end{figure}

Due to the non-monotonicity of the current as a function of $N$, the critical point
between the superconducting phase and the insulator phase can be better determined by
plotting the maximal supercurrent as a function of the ratio $E_J/E_C$ for different
circumferences.
The result is shown in Fig.~\ref{fig:cross-over} for the case $C=2C_0$.
%
%
%
%%%%%%%%%%%%%%%%%%%%%%%%%%%%%%%%%%%%%%%%%%%%%%%%%%%%%%%%%%%%%%%%%%%%%%%
%
%
%   FIG.12
%
%

We find that the critical value saturates to the value $E_J/E_0\simeq3.24$, as long as
$C \geq 2C_0$.
This result is in agreement with the result obtained by Korshunov\cite{Korshunov:1989}
who calculated  $E_J/E_0= 32/\pi^2$  in the regime $C \gg C_0$.
Reducing the ratio $C/C_0$, we find a small increase of the critical value $E_J/E_0$ which
saturates  to $\sim 4.16$ for vanishing mutual capacitance $C=0$.
In this limit, we do not recover the critical ratio $E_J/E_0= 128/\pi^2$ obtained by
Bradley and Doniach.\cite{BradleyDoniach:1984}

\subsection{Discussion}
\label{subsec:discussion}

In this section we discuss the validity of the single QPS approximation.

\paragraph{The case $C_0=0$.}
In this case,
it is most convenient to express the Lagrangian in terms of the
phase differences $\{ \theta_n \}$,
\begin{equation}
\label{eqn:Action_prova}
\mathcal{L} =
\sum_{n} \left[ \frac{\hbar^2 C}{8e^2}\dot{\theta}^2_n - E_J \cos\left( \theta_n +\frac{\delta}{N} \right)
\right]  \, ,
\end{equation}
showing that there is no correlation in the spatial direction.
Indeed, Eq.~(\ref{eqn:Action_prova})
with the constraint Eq.~(\ref{eq:constraint})
describes $N-1$ independent variables.
Under the condition $E_J>E_C$ one therefore can use the semiclassical instanton approach
to describe the QPS in each individual junction.
This is a well-controlled technique,\cite{Kleinert:1995}  based on the non-interacting
 (dilute) instanton approximation.
For instance, in the limit $N \gg 1$, the problem reduces to the tunneling of a free
particle in a double-well.
The resulting tunnel amplitude is $\nu_0$,
Eq.~(\ref{eq:v_0}), which is independent of the size of the system.
The total amplitude $N \nu_0$ grows linearly with N since
instantons which can occur independently in any of the junctions,
as previously analyzed in Ref.~\onlinecite{MateevLarkinGlazman:2002}.
The system becomes an insulator for $N \to \infty$; no phase transition occurs,
in agreement with conclusions obtained in the thermodynamic
limit.\cite{BradleyDoniach:1984,Korshunov:1986,Korshunov:1989}

\paragraph{The case  $C_0 \ll C$ and $N \alt \lambda$.}
When $C_0$ is restored, an interaction appears between the
phase-differences $\{ \theta_n \}$.\cite{Korshunov:1986}
This interaction  yields  a possible coupling between QPSs occurring in different junctions $n \neq m$.
This {\sl bare} interaction is proportional to $C_0/C$.
For finite systems and for $C_0 \ll C$ we expect that, by continuity with the case $C_0 = 0$,
this interaction can be neglected in first approximation.
Indeed, the instantons are still  rare events in the imaginary time for $E_J\gg E_C$ so that
one can study a single instanton centered in one junction $\theta_{n_0}$.
The other phase-differences $n \neq N-1$ can be approximated by $N-2$  harmonic
oscillators coupled to the winding phase $\theta_{n_0}$.

Another consequence of a finite $C_0$ is the fact that the
modes have a frequency dispersion.
This leads to an additional non-local term in the  effective action of $\theta_{n_0}$,
Eq.~(\ref{eqn:S_eff_C0}), beyond the finite-size corrections discussed for the case $C_0=0$.
In particular we considered the adiabatic regime in which all the modes have
a frequency higher than the tunneling frequency of the phase $\theta_{n_0}$.
This situation is very close to the case $C_0=0$:
the quantum tunneling of a fictitious particle in a
double-well can be still reduced to the tunneling between two levels but with an
adiabatically renormalized amplitude.
This is exactly the theoretical framework  presented in Ref.~\onlinecite{Leggett:1987},
in which the authors developed the analysis for a generic two-level system.
In particular they considered the adiabatic regime expressed by their Eq.~(2.9)
which corresponds to our Eq.~(\ref{eq:nu_general}).
We exploited this approach for a specific system, namely the
Josephson junction ring threaded  by a magnetic flux.
Moreover,  as shown in Sec.~\ref{sec:results_general} and in  Appendix \ref{app:analytic},
the corrections to the instanton action due to the non-local term are small as long as
$N \alt\lambda$.
Therefore, by continuity to the previous case, we expect that our results
are qualitatively and quantitatively correct in this regime.

\paragraph{The case $C_0 \ll C$ and $N \gg \lambda$.}
In the opposite limit $N \gg \lambda > 1$ (see Appendix \ref{app:analytic}), we find
 a logarithmic dependence on $N$  for the leading term in the instanton action
$S_{p}^{(cl)} \sim \pi \sqrt{E_J/(8E_0)} \log \left( N/\lambda \right)$.
Approaching the thermodynamic limit $(N \rightarrow \infty)$, even for $C_0 \ll C$, the
ground capacitance $C_0$ has a substantial effect.
It leads to a renormalized QPS amplitude which strongly depends on the length of the JJ ring
$\nu \sim \nu_0  \, N^{-\sqrt{\pi^2 E_J/(8 E_0)} }$.
%$\nu \sim \nu_0  \, N\!\!\wedge\!\!\{-\sqrt{\pi^2 E_J/(8 E_0)}\}$.

%
The validity of this result is addressed in the next subsection.

\subsection{The relation with the BKT transition.}
\label{subsec:discussion_BKT}

Mathematically, the one-dimensional quantum model of the type
in Eq.~(\ref{eq:Hamiltonian})  can be mapped onto a two-dimensional classical model.
Considering the axis $x =\tau$ (imaginary time) and the axis  $y=n$ (position on the chain)
{\sl for the local phases $\{\varphi_n(\tau)\}$}, Bradley and Doniach showed that, {\sl for the case $C=0$},
the 1D JJ chain is equivalent to an anisotropic 2D classical spins
with nearest neighbors interactions along the perpendicular axis.\cite{BradleyDoniach:1984}

Hence, according to this mapping, the superconductor-insulator phase transition
occurring in the case $N=\infty$ corresponds to the order-disorder phase transition
of an ensemble of ferromagnetically coupled planar classical spins.
This is the celebrated Berezinskii-Kosterlitz-Thouless (BKT) transition,
marked by a disruption of the ordered ferromagnetic phase due to the appearance of
vorticity: topological defects for which the spin orientation changes by $2\pi$ when
following a closed path around them once.
Reducing the ratio $E_J/E_0$, the BKT-transition is driven by the
dissociation of bound vortex -- anti-vortex pairs formed in the ordered phase for the
local phases  $\{\varphi_n(\tau)\}$.\cite{BradleyDoniach:1984}
Thus, as interaction between vortices plays an essential role in this scenario,
a natural question is the range of validity of our approach where the correlations between
QPSs have been ignored.

Korshunov in Ref.\onlinecite{Korshunov:1986} analyzed the same problem
$(C=0)$ but using the representation in the space of the phase-difference
$\{\theta_n(\tau)\}$.
In this space, the relevant (non-linear) quantum fluctuation are the instantons, {\em
i.e.}, the QPSs.
Notice that the vortices and the QPSs are not exactly  the same object: the former are
configurations defined in the space of  the local phase $\{\varphi_n(\tau)\}$ of the BCS
condensate in each superconducting island\cite{BradleyDoniach:1984} whereas the latter
are defined in the space of  the phase differences $\{\theta_n(\tau)\}$ in each Josephson
junction.\cite{Korshunov:1986}

The two instantons defined the $\theta$-space interact as\cite{Korshunov:1986}
\begin{equation}
\label{eq:S_interaction}
\mathcal{S}_{int}(\tau,\Delta N ) \sim \pi \sqrt{E_J/E_0} \log \left[
{\left(\omega_0 \tau \right)}^2
+ \Delta N^2 \right] \, ,
\end{equation}
in which $\omega_0\sim{(E_J E_0)}^{1/2}$ and  
$(\tau,\Delta N)$ are the separation along the imaginary time axis and
in real space, respectively.
Remarkably, two instantons interact again with a logarithmic potential in the two
directions in a very similar way to the vortices defined the $\varphi$-space.
This interaction leads to the formation of bound-pair states of QPSs which
makes the superconducting phase stable, even for an infinite number of QPSs.

This implies that, even if we focus on an individual junction $(\Delta N=0)$,
interaction appears between different instantons along the imaginary time.

The crucial point is that this interaction is mediated by the propagating modes on the
loop and is present only in the thermodynamic limit $N=\infty$, when the modes are
sufficiently dense.\cite{Leggett:1987}
In this limit, the local junction hosting a QPS is coupled to the rest of the infinite
chain which acts as a dissipative environment.
Indeed, the other Josephson junctions form a {\sl dense} bath of
harmonic oscillators  with linear low-frequency dispersion.
They mediate the interactions of two different instantons in imaginary time.
This is the reason for the appearance of non-local correlations in time between
different instantons.
As discussed  in Sec.~\ref{subsec:S_eff_N=infty}, the  ratio $C_0/C$  plays the role
of coupling between the winding phase and the harmonic bath of oscillators formed by
the other N-1 junctions.
Although the coupling constant is small $(C_0 \ll C)$, the dynamics of the winding phase is
now dissipative and the instantons in  imaginary time are strongly coupled  with a long-range
logarithmic potential:  the single QPS approximation breaks down.

{\sl As long as  $N$ remains finite, no real dissipation
appears and therefore no logarithmic interaction between instantons occurs in imaginary
time.}
In particular we studied the adiabatic regime in which the problem  can be
reduced to the renormalized  quantum tunneling of a fictitious particle in a
double-well potential\cite{Leggett:1987} (see the cases {\sl b} and {\sl c} in
Sec.~\ref{subsec:discussion}).

Between these two limits, we have a continuous cross-over from the adiabatic regime
to the full dissipative dynamics at $N=\infty$.
Therefore we expect that, at given ratio $C/C_0$ and $E_J/E_C$, there is a typical length
$N^*$ which sets an upper bound for the range of validity of the single QPS
approximation.

Indeed, increasing the length $N$, the lowest frequency $\omega_{min}$ defined in
Eq.~(\ref{eq:om_min})  decreases as $\hbar \omega_{min} \sim 4\sqrt{2 E_J E_C}(\lambda/N)$.
Then, for our approach to be self-consistent, we estimate $N^*$ as the point at which the
adiabatic condition  breaks down, namely when the level splitting of the two level system
coincides with the lowest frequency of the modes: $2 \nu_0 \sim \hbar\omega_{min}$.
It gives
\begin{equation}
\label{eq:N^*} N^*/\lambda =  (\sqrt{\pi}/2^{3/4}) \, e^{\sqrt{8\frac{E_J}{E_C}}} /
{(E_J/E_C)}^{1/4} \, .
\end{equation}
As an example, at $E_J/E_C= 6.5$,  $\lambda=4.4$, $(C=2C_0)$ (the values close to the
phase-transition, see Fig.~\ref{fig:cross-over}) we have the condition $N^* \sim 2800$.

For $N \agt N^*$ only a finite number of discrete  modes have frequencies lower than the
tunneling frequency. 
\footnote{Notice that, when we consider the regime $C_0>0$ and $N\gg\lambda$, 
the condition of validity $N<N^*$ also implies an upper bound 
to the parameter $q$ defined in Eq.~(\ref{eq:q-factor}).}
Notice that this  region beyond $N \agt N^*$ is still far  away from
the full dissipative limit $N=\infty$, in which we expect a logarithmic interaction
between instantons Eq.~(\ref{eq:S_interaction}). Therefore one has to consider
Eq.~(\ref{eq:N^*}) as a rough estimate. Remarkably,  as shown in
Fig.~\ref{fig:cross-over}, in finite-size systems with $N \sim 10^3$ and well below the
line $N=\infty$, the crossing point of the maximal current can be very close to the
critical value of the phase-transition although the current does not drop vertically to
zero as expected in the real thermodynamic limit.

We conclude by recalling that the exact analysis of the interaction between two
instantons in  imaginary time beyond the limit $N \gg N^*$ in JJ chains is an interesting
and open theoretical issue, beyond the aim  of  the present work. Moreover we remark that
the Lagrangian of Eq.~(\ref{eqn:Action_phi_n}) can be exactly mapped on the classical
(anisotropic) XY spin-Hamiltonian -- which represents the reference model for the BKT
transition -- {\sl only}  for $C=0$.
For the general case of $(C,C_0)$, the two-dimensional classical spin
Hamiltonian associated with the Lagrangian of Eq.~(\ref{eqn:Action_phi_n}) has a
four-body anisotropic-diagonal interaction between the spins $\varphi_n(\tau)$.
Even at $N=\infty$, the superconductor/insulator transition in
the extreme regime $C \gg C_0$  has not yet been analyzed in detail in the literature
(with the exception of Ref.~\onlinecite{Korshunov:1989}).

\section{Summary and Conclusions}
\label{sec:conclusions}

In this work we studied the quantum QPS processes in 1D Josephson junction rings
in the strong Josephson coupling limit $E_J \gg E_C,E_0$.
In contrast with the previous work,\cite{MateevLarkinGlazman:2002}
we consider JJ rings of finite size in a wide range of lengths $( N \agt 5)$
and with ground capacitance $C_0$.
We calculated  the renormalized QPS amplitude $\nu(N)$ and we discussed its consequence for
the maximum supercurrent $I_{max}$ flowing through a JJ ring threaded by a magnetic
flux.

For the case $C_0=0$, we found an interplay between different finite-size effects
which gives rise to a non-monotonic  behavior of the maximum supercurrent
$I_{max}/I_\mathrm{cl}$ as a function of the ring size.
The critical length above which finite size corrections are negligible is
$N_c \sim (\pi^2 + 4) \sqrt{8E_J/E_C}$.

When the ground capacitance is restored, $C_0>0$, dispersive modes are possible
on the ring which are directly coupled to the local winding phase-difference.
When $N \rightarrow \infty$  we found that the system  converges either
to a superconducting state with $I_{max}/I_\mathrm{cl} = 1$ or to an insulating phase
$I_{max}/I_\mathrm{cl} = 0$, depending on the ratio $E_J/E_0$.
For $C > C_0$ we found as critical ratio ${(E_J/E_0)}_c=3.24$,
in agreement with the previous work of Ref.~\onlinecite{Korshunov:1989}.

Although our analysis was mainly developed for a ring of identical junctions, it is also
relevant for other systems, for instance for a weak winding junction coupled to a chain
of Josephson junctions in which the QPS is prevented, {\em i.e.}, the so-called
fluxonium.\cite{Manucharyan:2009}

We have discussed the validity of the single QPS approximation.
The regime $E_C > E_J$ associated with the regime of strong interaction
between the QPSs is beyond the scope of this article.
A global phase diagram was reported some time ago in Refs.~\onlinecite{Korshunov:1986,Korshunov:1989}
in the thermodynamic limit but it was based on a perturbative renormalization group analysis.
A phase-diagram obtained by non-perturbative approaches
as well as the behavior of finite-size systems for arbitrary
ranges of the parameters $E_J,E_0,E_C$ and $C/C_0$ constitutes
a still open theoretical issue.

\acknowledgments 
We thank W.\ Guichard, L.\ Glazman, M.\ Vanevi{\'c}, L.\ Amico and L.\ Ioffe for useful discussions. 
This work was supported by ANR through contracts DYCOSMA and QUANTJO. 
We acknowledge support from the European networks MIDAS, SOLID and GEOMDISS 
and from Institut universitaire de France.

\appendix

\section{Path integral}
\label{app:integration}
In this Appendix we summarize the main steps for the calculation of
the path integral Eq.~(\ref{eq:Z_tot_CL}).

First we recall that any periodic function $\varphi_n$ on the lattice $n=0,\dots,N-1$ can
be decomposed as Eq.~(\ref{eq:eigenmodes}) where the set $\{ \varphi_k \}$ are complex
numbers, {\em i.e.}, $\varphi_k = \varphi_k^R + i \varphi_k^I$.
They are related by the condition $\varphi_{N-k} = \varphi_{k}^*$
as $\varphi_n$ is real.
Using this property, we can write
\begin{equation}
\label{eq:varphi_n}
\!\varphi_n\!\!= \!
\frac{1}{\sqrt{N}}\!\!
\left[
\varphi_{k=0} + \underbrace{{(-1)}^{n} \varphi_{k=\frac{N}{2}}}_{\mbox{\small N even}}
+
\!\!\!
\sum_{k=1}^{k_{max}}
\left(\!
\varphi_k e^{i\frac{2\pi k}{N}n} + c.c.\!
\right)
\right] .
\end{equation}
For $N$ even we have $k_{max}= (N/2)-1$
whereas for $N$ odd we have $k_{max}= (N-1)/2$.
We can express the Euclidean Lagrangians only in terms of
the independent variables.
Setting the action
\begin{equation}
\label{eq:L_harmonic}
L_0\left(x_k\right) =
\frac{1}{2} \mu_k \dot{x}^2_k +
\frac{1}{2} \mu_k \omega_k^2 x^2_k \, ,
\end{equation}
one can demonstrate that the harmonic Lagrangian  Eq.~(\ref{eqn:Action_phi_n_harm})
(omitting the constant proportional to $\delta_m$) in terms of the modes $k$
reads
\begin{equation}
L^{(N)}_0
=
\frac{\mu_0}{2} \dot{\varphi}^2_{k=0}
+
\underbrace{L_0(\varphi_{k=\frac{N}{2}})}_{\mbox{\small N even}}
+
2 \! \sum_{k=1}^{k_{max}}
\left[
L_0 \!\left( \varphi_{k}^R \right)
+
L_0 \!\left( \varphi_{k}^I \right)
\right] .
\end{equation}

We now consider the actions Eqs.~(\ref{eq:L_1ps}), (\ref{eq:L_bath}), and  (\ref{eq:L_int})
where we have $N-1$ harmonic variables $\varphi_n$.

Inserting Eq.~(\ref{eq:varphi_n}) with $N$ replaced by $N-1$ in Eq.~(\ref{eq:L_int}), one
can obtain  that only the imaginary parts of the  modes are linearly coupled to external
forces $(\theta(\tau),\dot{\theta}(\tau))$ through the position and through the velocity,
{\em i.e.},  Eq.~(\ref{eq:L_int_diag}).
For the notation we set
\begin{equation}
q_k(\tau) \equiv \varphi_k^{I}(\tau)  \, ,
\end{equation}
and we can write
\begin{equation}
\label{eq:_L_3_app}
\mathcal{L}_{3}= \sum_{k=1}^{k_{max}} \zeta_k
\left(
\frac{\hbar^2 C}{8e^2} \dot{q}_k \dot{\theta} +
\frac{E_J}{2} q_k \theta \right) \, .
\end{equation}
From Eq.~(\ref{eq:varphi_n}) we observe that the phase $\varphi_{n=0}$ at $n=0$
depends only on the real part of the mode $\{\varphi_k^R\}$.
Therefore the first term of Eq.~(\ref{eq:L_bath}), $\varphi^2_{n=0}$,  couples only the real parts
of the different modes $k$.
As real and imaginary part are decoupled,  the relevant term in Eq.~(\ref{eq:L_bath_diag})
in which we are interested for the calculation of the effective action reduces to
\begin{equation}
\label{eq:_L_2_app}
\mathcal{L}_{2}
\sim
\mathcal{E}_0
+
\!
2 \sum_{k=1}^{k_{max}}
\left( \frac{1}{2} \mu_k \dot{q}^2_k +
\frac{1}{2} \mu_k \omega_k^2 q^2_k
\right)
\, ,
\end{equation}
with $\mathcal{E}_0 = E_J\delta_m^2(N-1)/(2N^2)$.
We express the generic periodic path of the partition function as
\begin{equation}
\label{eq:phi_tau}
q_{k}(\tau) = q_{k,0} + \sum^{+\infty}_{\ell=1}
\left( q_{k,\ell} e^{i \omega_{\ell} \tau } + c.c. \right) \, ,
\end{equation}
in which the Fourier (Matsubara) component at $\omega_{\ell}=2\pi\ell/\beta$ is given by
$q_{k,\ell} = (1/\beta)\int^{\beta}_0 \!\! d\tau  \exp(- i \omega_{\ell} \tau) q_{k}(\tau) $.
Inserting the expression Eq.~(\ref{eq:phi_tau}) into the
Lagrangian  $\mathcal{L}_2+\mathcal{L}_3$ Eqs.~(\ref{eq:_L_3_app}),(\ref{eq:_L_2_app})
and integrating over the time, we obtain
\begin{equation}
\int^{\beta}_0 \!\!\!\! d\tau \left( \mathcal{L}_2 + \mathcal{L}_3 \right) =
\mathcal{E}_0
+
2\!\sum_{k=1}^{k_{max}} \left( \mathcal{S}_{k,0}  +
\sum_{\ell=1}^{+\infty}
 \mathcal{S}_{k,\ell} \right) \, .
\end{equation}
The first term $\mathcal{S}_{k,0}$ contains only the component at zero frequency
of $\theta(\tau)$ and $q_k(\tau)$,
\begin{equation}
\mathcal{S}_{k,0} =
\frac{\beta}{2} \mu_k \omega_k^2 q^2_{k,0}
+
\frac{\beta E_J}{2} q_{k,0} \theta_{0}
\, .
\end{equation}
The second terms $\mathcal{S}_{k,\ell}$ contain all the nonzero
frequency components $(\ell>0)$
\begin{equation}
\mathcal{S}_{k,\ell} \!\! =\!
\beta
\mu_k \left( \omega_{\ell}^2  + \omega_k^2 \right)  {\left| q_{k,l} \right|}^2
\!
+
\frac{\beta \zeta_k\hbar^2}{16 E_C}
(\omega_{\ell}^2  + \omega_p^2)
\left(q_{k,\ell}^* \theta_{\ell} + c.c.\right)
\end{equation}
Finally, the path integral is evaluated by integration over the Fourier
components as
\begin{eqnarray}
\Delta \mathcal{Z} &=& \!\!\!  \prod^{k_{max}}_{k=1} \!
\oint \!\! \mathcal{D} \left[ q_k(\tau)\right] \,
e^{-\frac{1}{\hbar} \int^{\beta}_0 \!\!\! d \tau
\left(\mathcal{L}_{2}+ \mathcal{L}_{3}  \right) }
= e^{-\frac{\beta \mathcal{E}_0}{\hbar}}  \nonumber \\
& \times &
\!\!\!\!\!
\prod^{k_{max}}_{k=1}
\!\!
\int \!\!
\frac{dq_{k,0} \, e^{-\frac{2 \mathcal{S}_{k,0}}{\hbar}} }
{\sqrt{\frac{4\pi\beta\hbar}{\mu_k}}}
\prod^{+\infty}_{\ell=1}
\iint \!\!
\frac{dq_{k,\ell}^{R}dq_{k,\ell}^{I} }{\frac{\pi \hbar}{\beta \mu_k \omega_{\ell}^2}}
\,\,  e^{-\frac{2\mathcal{S}_{k,\ell}}{\hbar}}
\label{eq:Delta_Z_calculus}  \, , \nonumber \\
& &
\end{eqnarray}
where the pre-exponential factors are the Jacobians associated
with the transformation of the path integral from the time-space to the
frequency-space.\cite{Kleinert:1995}
$(q_{k,\ell}^{R},q_{k,\ell}^{I})$ denote  respectively
the real and imaginary part of $q_{k,\ell}$.
The integral Eq.~(\ref{eq:Delta_Z_calculus}) has the general Gaussian form and
its evaluation is straightforward.
After some algebra, the relevant exponential term is
\begin{equation}
\label{eq:Delta_Z_calculus_2}
\Delta \mathcal{Z} \sim
\exp
\left\{
\frac{\beta\hbar}{8 E_C}
\sum_{\ell=0}^{\infty} Y(\omega_{\ell})
\frac{(\omega_{\ell}^2+\omega_p^2)}{1+\delta_{\ell,0}}
{\left| \theta_{\ell} \right|}^2
\right\}
 \, ,
\end{equation}
with
\begin{equation}
\label{eq:Delta_Z_calculus_3}
 Y(\omega_{\ell}) =
\frac{1}{N-1}
\sum_{k=1}^{k_{max}}
\frac{\sin^2\left(\frac{2\pi k}{N-1} \right)}
{1\!\!-\!\!\cos\left(\! \frac{2\pi k}{N-1} \! \right) \!\!+ \!\!
\frac{\pi^2}{2\lambda^2} \left( \frac{\omega_{\ell}^2}{\omega_{\ell}^2+\omega_{p}^2} \right) }
\, .
\end{equation}
By adding the exponential term of Eqs.~(\ref{eq:Delta_Z_calculus_2}),(\ref{eq:Delta_Z_calculus_3})
to the Lagrangian  $\mathcal{L}_1$ Eq.~(\ref{eq:L_1ps}) and going back to the time representation,
we obtain the result shown in Eqs.~(\ref{eq:Z_final}), (\ref{eqn:S_eff}),
and (\ref{eq:G_ell}).

\section{Instanton solution in periodic parabolic potential}
\label{app:parabolic}

Referring to the action Eq.~(\ref{eq:S_eff_parabol}), we set
$\hbar \omega_J = 4V_J e^2/(C^*+C_0/2)$ as a short-hand notation.
The most general path can be always expressed as
\begin{equation}
\label{eq:theta_tau_1}
\theta\left(\tau\right) = \theta\left(  -\beta/2 \right)
+ \int_{-\frac{\beta}{2}}^{\tau}\!\!\!\!\!\! d\tau' \: \dot{\theta}\left(\tau'\right)
\, .
\end{equation}
For the asymptotic instanton-like solution, we require the following
boundary  conditions at the end-points
$\theta_i= \theta( -\beta/2 )$ and $\theta_f= \theta( \beta/2 )$
\begin{equation}
\label{eq:theta_tau_2}
\lim_{\beta\rightarrow+\infty}  \theta_i = 0 \, , \,
\lim_{\beta\rightarrow+\infty}  \theta_f = \Delta \theta^* \, .
\end{equation}
It is useful to use the Fourier components of the velocity as free variables
\begin{equation}
\label{eq:theta_tau_3}
\dot{\theta}\left(\tau\right) = \sum^{+\infty}_{\ell=-\infty}
\dot{\theta}_{\ell} \, e^{i \omega_{\ell} \tau}
\, , \quad
\dot{\theta}_{\ell} = \frac{1}{\beta}
\!\int^{\frac{\beta}{2}}_{-\frac{\beta}{2}}\!\!\!\!\! d\tau \,
 \dot{\theta}\left(\tau\right) e^{-i \omega_{\ell} \tau}  \, .
\end{equation}
By assuming that the total energy is conserved,
we impose that the initial velocity is equal to the final one:
$\dot{\theta}(-\beta/2)=\dot{\theta}(\beta/2)$.
More specifically,
due to the symmetry of the potential, we can assume that the velocity is
an even function of the time $\dot{\theta}(\tau)= \dot{\theta}_{0}
+\sum^{+\infty}_{\ell=1} 2 \cos( \omega_{\ell} \tau ) \dot{\theta}_{\ell} $,
with $\dot{\theta}_{\ell}$ are real numbers.
By definition, the average velocity is $\dot{\theta}_{0} = (\theta_f-\theta_i)/ \beta$.
In terms of the variables $\{ \dot{\theta}_{\ell} \}$, the path Eq.~(\ref{eq:theta_tau_1})
reads
\begin{equation}
\label{eq:theta_tau_5}
\theta\left(\tau\right) =
\theta_i +
\dot{\theta}_{0} \left( \tau + \beta/2 \right) + \sum_{\ell\neq0}
\left( \frac{e^{i \omega_{\ell} \tau}
- e^{-i \omega_{\ell} \frac{\beta}{2}}}{i \omega_{\ell}} \right)
\dot{\theta}_{\ell}
\, .
\end{equation}
We recall the definition of the kernel $G(\tau)$ in the action
Eq.~(\ref{eq:S_eff_parabol}) in the Fourier space
\begin{equation}
\label{eq:theta_tau_6}
G\left(\tau\right) =  \frac{1}{\beta} \sum_{\ell \neq 0} G_{\ell} \,
e^{i \omega_{\ell} \tau}  \, , \quad \mbox{with} \,\, G_{-\ell} =G^{*}_{\ell} \, .
\end{equation}
Using the expressions
Eqs.~(\ref{eq:theta_tau_3}),(\ref{eq:theta_tau_5}),(\ref{eq:theta_tau_6})
in the action Eq.~(\ref{eq:S_eff_parabol}) and taking into account the symmetry
respect to the time, the time integration is straightforward.
The general action is thus expressed in term of the variables~$\{ \dot{\theta}_{\ell} \}$
\begin{small}
\begin{eqnarray}
\frac{1}{\beta}S_{p}\left( \left\{ \theta_{\ell} \right\} \right) &=&
\frac{V_J}{2\omega_J^2} \dot{\theta}_0^2
+
\sum_{\ell=1}^{+\infty}
\left( \frac{V_J}{\omega_J^2} + \frac{V_J}{\omega_{\ell}^2} +  \frac{G_{\ell}}{\omega_{\ell}^2}  \right)
\dot{\theta}^2_{\ell}  \nonumber \\
&-&
\sum_{\ell=1}^{+\infty}
\frac{4 V_J}{\beta \omega_{\ell}^2}
\left[
 \theta_i \left( 1-(-1)^{\ell} \right)
+ \frac{\beta \dot{\theta}_0}{2}\right] \dot{\theta}_{\ell} \nonumber \\
&+& \frac{V_J}{2} \left[ \theta_i^2 + \frac{\beta
\dot{\theta}_0}{2} \theta_i + \frac{ {\left( \beta
\dot{\theta}_0 \right)}^2}{12} \right] \!\! \label{eqn:S_par_theta_ell}.
\end{eqnarray}
\end{small}
Then, by the condition of minimization $\partial S_{p}/\partial \dot{\theta}_{\ell} = 0$,
we find the classical solution $\dot{\theta}_{\ell}^{(cl)}$
\begin{equation}
\label{eq:theta_solution}
\dot{\theta}_{\ell}^{(cl)} =
\frac{2\omega^2_J/\beta}{\omega^2_{\ell}+\omega^2_J+  \frac{G_{\ell}}{V_J} \omega^2_J }
\left[
\left[ 1-(-1)^{\ell} \right] \theta_i  +
\frac{\beta \dot{\theta}_0}{2}
\right] \, .
\end{equation}
We insert the solution Eq.~(\ref{eq:theta_solution}) into the action
Eq.~(\ref{eqn:S_par_theta_ell}) to obtain the value of the action at the
minimum
\begin{equation}
\frac{ S^{(cl)}_{p} }{\hbar} =
\frac{\beta V_J\dot{\theta}_0^2}{2\hbar\omega_J^2}
+ \frac{4 V_J}{\hbar\beta}
\sum_{\ell=1}^{+\infty}
\frac{
{\left[
\left[ 1-(-1)^{\ell} \right] \theta_i   +
\frac{\beta \dot{\theta}_0}{2}
\right]}^2
}{\omega_{\ell}^2+ \frac{\omega^2_J}{1+
\omega_J^2 G_{\ell}/(\omega_{\ell}^2 V_J)} }  \, .
\end{equation}
Finally, we take the limit $\beta \rightarrow \infty$.
We use the boundary conditions Eq.~(\ref{eq:theta_tau_2})
and the fact that the average velocity $\dot{\theta}_{0}$ scales as
$(\theta_f-\theta_i)/\beta \simeq \Delta\theta^*/\beta$.
After that, the resulting series with respect to Matsubara frequencies
$\ell=1,\dots,\infty$ converges to an integral $\omega_{\ell}=\omega$
and the resulting action coincides with Eq.~(\ref{eq:S_par}) of the main text
by recalling $\hbar \omega_J = 4V_J e^2/(C^*+C_0/2)$ and
Eq.~(\ref{eq:theta_ren}) for $\Delta\theta^{*}$.

\section{Long JJ ring with ground capacitance}
\label{app:analytic}

In this Appendix  we discuss some general features and some analytic limits of
the effective action obtained within the parabolic approximation Eq.~(\ref{eq:S_par}).
To simplify the notation we set $V_J = E_J$ as the results are
qualitatively the same for two different coefficients.
We focus on the {\sl long circumferences limit} of the rings defined by the condition that we can
neglect the corrections of order $(1/N)$ in $E_C^*$, $E_J^*$ and $\Delta\theta^*$ as well
as $N \gg \max(\lambda,1)$.
In this regime, we replace the sum with respect to the modes $k$ in
$G(\omega_{\ell}) \equiv G(\omega)$ Eq.~(\ref{eq:G_ell}) with an integral
and we obtain
\begin{equation}
\label{eq:G_omega_asy}
G\left(\omega\right)  = \frac{{(\hbar\omega)}^2}{16e^2} C_0
\left( F\left(\omega\right)  \sqrt{1+\frac{4C}{C_0}} - 1 \right)
\, ,
\end{equation}
where we set
\begin{equation}
\label{eq:F_omega}
F\left(\omega\right) = \sqrt{1 + \frac{\omega_{max}^2}{\omega^2}}
\left[
\! 1 \! - \!  \frac{2}{\pi} \arctan\left( \frac{\omega_{min}}{\omega}
\sqrt{1 + \frac{\omega_{max}^2}{\omega^2}}
\right)
\!  \right]
\, ,
\end{equation}
where $\omega_{max},\omega_{min}$ are defined in Eqs.~(\ref{eq:om_max}),(\ref{eq:om_min}).
Note that we can not neglect the $N-$dependence in the function $F(\omega)$,
Eq.~(\ref{eq:F_omega}), as it is strongly dependent on the minimum cut-off
frequency $\omega_{min}$.
Using the Eqs.~(\ref{eq:G_omega_asy}),(\ref{eq:F_omega}), we can express
the action on the classical instanton path Eq.~(\ref{eq:S_par}) as
\begin{equation}
\label{eq:S_cl_p_asym}
\frac{S^{(cl)}_p}{\hbar}=
2\pi \frac{E_J}{\hbar}
\int^{\infty}_{0}  \!\!\!\!\!\!\! d\omega
\frac{1}{\omega^2+
\frac{\omega^2_{max}}{1+ 2F(\omega)/{(1+4C/C_0)}^{1/2}}}
\, .
\end{equation}
The two frequencies $\omega_{max},\omega_{min}$ define three ranges for the integral
on the frequency $\omega$.
They are: i) $\omega_{max} \ll \omega$,
ii) $\omega_{min} \ll  \omega \ll  \omega_{max} $, and  iii) $\omega \ll \omega_{min}$.
In these range the function $F(\omega)$ can be approximated as: i)
$F(\omega) \simeq 1$, ii) $F(\omega) \simeq \omega_{max}/\omega$, and
iii) $F(\omega) \simeq (N/\pi)/{(1+4C/C_0)}^{1/2}$, to leading order in $1/N$.
Cutting the integral Eq.~(\ref{eq:S_cl_p_asym}) in three parts,
we use the three approximated expression for the function $F(\omega)$
to evaluate the integration
\begin{equation}
\int^{+\infty}_0 \!\!\!\!\!\!\!\! d\omega \dots =
\int^{+\infty}_{\omega_{max}} \!\!\!\!\!\!\!\! d\omega \dots
+ \int^{\omega_{max}}_{\omega_{min}} \!\!\!\!\!\!\!\! d\omega \dots
+ \int^{\omega_{min}}_0 \!\!\!\!\!\!\!\! d\omega \dots \,\,\, .
\end{equation}
The first integral  i) in the high-frequency range as well as
the third  integral iii) in the low-frequency range gives a result
independent of the ring's size $N$.
On the other hand, for the second integral ii) in the intermediate
frequency range we find the important result
\begin{eqnarray}
\frac{S^{(cl)}_p}{\hbar} &\sim&  2\pi \frac{E_J}{\hbar}
\int^{\omega_{max}}_{\omega_{min}}  \!\!\!\!\!\!\!\!\!\!\!\!\! d\omega \,\,
\frac{1}{\omega^2+  \omega \, \omega_{max} \, {(1+4C/C_0)}^{1/2}} \nonumber \\
&\sim& \pi  \sqrt{\frac{E_J}{8E_0}} \log\left( \frac{N}{\lambda} \right) + \dots \, .
\end{eqnarray}
As we have reported numerically, we have a logarithmic dependence on $N$ of the
classical action yielding to a power-law dependence of the
QPS amplitude $\nu$.
This leads  to a superconductor/insulator phase transition when the
limit $N=\infty$  is taken.

Finally, we observe that the function $F(\omega)$ Eq.~(\ref{eq:F_omega})
saturates to a constant for $\omega \ll \omega_{min}$
This low-frequency cut-off is important because it makes that the total
integral convergent.
As we have explained in the text, below this threshold $\omega \ll \omega_{min} $,
the winding junction feels the discreteness of the spectrum of the environment
with which it can exchange energy.
That corresponds to say that there is no real dissipation at low frequency and an
instanton solution conserving the initial energy still exists.


\begin{thebibliography}{99}

%
%   MOTIVATIONS
%

\bibitem{Ioffe:2002Nature}
L.\ B.\ Ioffe, M.\ V.\ Feigel'man, A.\ Ioselevich, D.\ Ivanov, M.\ Troyer, and G.\ Blatter,
Nature {\bf 415}, 503 (2002).

\bibitem{Ioffe:2002}
L.\ B.\ Ioffe and M.\ V.\ Feigel'man, Phys.\ Rev.\ B {\bf 66}, 224503 (2002).

\bibitem{Douçot:2002}
B. Dou\c{c}ot and J. Vidal, Phys.\ Rev.\ Lett.\ {\bf 88}, 227005 (2002).

\bibitem{Douçot:2003}
B.\ Dou\c{c}ot, M.\ V.\ Feigel'man, and L.\ B.\ Ioffe, Phys.\ Rev.\ Lett.\ {\bf 90}, 107003 (2003).

\bibitem{Douçot:2005}
B.\ Dou\c{c}ot, M.\ V.\ Feigel'man, L.\ B.\ Ioffe, and A.\ S.\ Ioselevich, Phys.\ Rev.\ B {\bf 71}, 024505 (2005).

\bibitem{Gladchenko:2009}
S.\ Gladchenko, D.\ Olaya, E.\ Dupont-Ferrier, B.\ Dou\c{c}ot, L.\ B.\ Ioffe, and
M.\ E.\ Gershenson, Nat. Physics {\bf 5}, 48 (2009).

\bibitem{Castellanos-Beltrana:2007}
M.\ A.\ Castellanos-Beltrana and  K.\ W.\ Lehnert, Applied Physics Letters {\bf 91},
083509 (2007).

\bibitem{Castellanos-Beltrana:2008}
M.\ A.\ Castellanos-Beltrana, K.\ D.\ Irwin, G.\ C.\ Hilton, and  L.\ R.\ Vale,
Nat. Physics {\bf 4}, 928 (2008).

\bibitem{GuichardHekking:2010}
W.\ Guichard and F.\ W.\ J.\ Hekking, Phys.\ Rev.\ B {\bf 81}, 064508 (2010).

\bibitem{Flowers:2004}
J.\ Flowers, Science {\bf 306}, 1324 (2004).

\bibitem{Manucharyan:2009}
V.\ E.\ Manucharyan, J.\ Koch, L.\ I.\ Glazman, and M.\ H.\ Devoret, Science {\bf 326}, 113 (2009).

\bibitem{Haviland:2011}
C.\ Hutter, E.\ A.\ Thol{\'e}n, K.\ Stannigel, J.\ Lidmar, and D.\ B.\ Haviland, Phys.\ Rev.\ B {\bf 83}, 014511 (2011).

%
%   THEORY
%

\bibitem{BradleyDoniach:1984}
R.\ M.\ Bradley and S.\ Doniach, Phys.\ Rev.\ B {\bf 30}, 1138 (1984).

\bibitem{Korshunov:1986}
S.\ E.\ Korshunov, Sov.\ Phys.\ JETP {\bf 63}, 1242 (1986).

\bibitem{Korshunov:1989}
S.\ E.\ Korshunov, Sov.\ Phys.\ JETP {\bf 68}, 609 (1989).

\bibitem{Fazio:2001}
R.\ Fazio and H. van der Zant, Phys.\ Rep.\ {\bf 355}, 235 (2001).

\bibitem{Sarkar:2007}
S. Sarkar, Phys.\ Rev.\ B {\bf 75}, 014528 (2007).

\bibitem{Sarkar:2009}
S. Sarkar, Eur.\ Phys.\ J.\ B {\bf 67}, 559 (2009).


%
%   EXPERIMENTS
%

\bibitem{Chow:1998}
E.\ Chow, P.\ Delsing, and D.\ B.\ Haviland, Phys.\ Rev.\ Lett.\ {\bf 81}, 204 (1998).

\bibitem{Haviland:2000}
D.\ B.\ Haviland, K.\ Andersson, and P.\ {\AA}gren, Jour. of Low Temp. Physics
{\bf 118}, 733 (2000).

\bibitem{Kuo:2001}
W.\ Kuo and C.\ D.\ Chen, Phys.\ Rev.\ Lett.\ {\bf 87}, 186804 (2001).

\bibitem{Miyazaki:2002}
H.\ Miyazaki, Y.\ Takahide, A.\ Kanda, and Y.\ Ootuka, Phys.\ Rev.\ Lett.\ {\bf 89}, 197001
(2002).

\bibitem{Takahide:2006}
 Y.\ Takahide, H.\ Miyazaki, and Y.\ Ootuka, Phys.\ Rev.\ B {\bf 73}, 224503 (2006).


%
%   Recent works + others
%


\bibitem{MateevLarkinGlazman:2002}
K.\ A.\ Matveev,  A.\ I.\ Larkin, and L.\ I.\ Glazman, Phys.\ Rev.\ Lett.\ {\bf 89}, 096802 (2002).

\bibitem{Pop:2010}
I.\ M.\ Pop, I.\ Protopopov, F.\ Lecocq, Z.\ Peng, B.\ Pannetier, O.\ Buisson,  and W.\ Guichard,
Nat. Physics {\bf 6}, 589 (2010).

\bibitem{Orlando:1999}
T.\ P.\ Orlando, J.\ E.\ Mooij, L.\ Tian, C.\ H.\ van der Wal, L.\ S.\ Levitov, S.\ Lyod,
J.\ J.\ Mazo, Phys.\ Reb.\ B {\bf 60}, 15398 (1999).


\bibitem{Likharev:1985}
K.\ K.\ Likharev and A.\ B.\ Zorin,  J.\ Low Temp.\ Phys.\ {\bf 59}, 347 (1985).

\bibitem{Catelani:2011}
G.\ Catelani, R.\ J.\ Schoelkopf, M.\ H.\ Devoret, and L.\ I.\ Glazman, Phys. Rev. B {\bf 84},
064517 (2011).

\bibitem{note:factor_2}
Note that Eq.~(\ref{eq:v_0}) corresponds to $\bar{\epsilon}/2$ in Eq.~100 of
Ref.\onlinecite{Catelani:2011}.

\bibitem{Masluk:2012}
N.\ A.\ Masluk, I.\ M.\ Pop, A.\ Kamal, Z.\ K.\ Minev, and M.\ H.\ Devoret, Phys.\ Rev.\ Lett.\ {\bf 109}, 137002 (2012).

\bibitem{Caldeira:1981}
A.\ O.\ Caldeira and A.\ J.\ Leggett, Phys.\ Rev.\ Lett.\ {\bf 46}, 211 (1981).

\bibitem{Schmid:1983}
A.\ Schmid, Phys.\ Rev.\ Lett.\ {\bf 51}, 1506 (1983).

\bibitem{note:charge-solitons}
Note that the model discussed here is different from the one introduced by Z.\ Hermon,
E.\ Ben-Jacob, and G.\ Sch{\"o}n, Phys.\ Rev.\ B {\bf 54}, 1234 (1996) in which the kinetic
inductance of the superconducting grains was assumed to be much  larger  than the
Josephson inductance. Here we consider the  experimentally  more relevant opposite
regime where the Josephson inductance dominates (see also Ref.~\onlinecite{Hekking:1997}).

\bibitem{Hekking:1997}
F.\ W.\ J.\ Hekking and  L.\ I.\ Glazman, Phys.\ Rev.\ B {\bf 55}, 6551 (1997).

\bibitem{Tinkham:1996}
M.\ Tinkham, {\em Introduction to Superconductivity} (McGraw-Hill int. eds., Singapore
1996, 2nd edition).

\bibitem{Pop:2011b}
I.\ M.\ Pop, B.\ Dou\c{c}ot, L.\ Ioffe, I.\ Protopopov, F.\ Lecocq, I.\ Matei,
O.\ Buisson, and W. Guichard, Phys. Rev. B {\bf 85}, 094503 (2012).

%
%
%
%   TECHNICAL PAPERS
%
%
%

\bibitem{footnote:detailed-comparison}
The detailed comparison of our result with Eq.~(8) of
Ref.~\onlinecite{Korshunov:1989}
requires the substitution $E_J \rightarrow \hbar V$ and $\hbar^2C/(4e^2)  \rightarrow M,
\hbar^2C_0/(4e^2)\rightarrow m$.


\bibitem{Schon:1990}
G.\ Sch{\"o}n and A.D. Zaikin,  Phys. Rep.  {\bf 198},  237 (1990).

\bibitem{Rastelli:2012}
G.\ Rastelli, Phys.\ Rev.\ A {\bf 86}, 012106 (2012).

\bibitem{asymmetric-double-well}
P.\ R.\ Johnson et al., Phys.\ Rev.\ Letts.\ {\bf 94}, 187004 (2005);
J.\ M.\ Schmidt, A.\ N.\ Cleland, and J.\ Clarke, Phys.\ Rev.\ B {\bf 43}, 229 (1991).

\bibitem{Kleinert:1995}
H.\ Kleinert, {\em Path Integral in Quantum Mechanics, Statistics and Polymer Physics.}
(World Scientific, Singapore 1995, 2nd edition).

\bibitem{Villain:1975}
J.\ Villain, J.\ Phys.\ (Paris) {\bf 36}, 581 (1975).

\bibitem{note:parabolic_approximation}
Note that for $G(\omega)=0$  $(C_0=0)$, the integral converges
to the factor $\pi/2$ and we obtain
$S_{par}/\hbar = 8 {(\Delta \theta^*/(2\pi))}^2 \sqrt{E_J^*/E_C^*}$.
Actually, this result is slighlty different from the
one obtained with the cosine potential Eq.~(\ref{eq:v_ren}).
In the latter case, the classical action depends linearly
on $\Delta \theta^*/(2\pi) = 1-1/N$.
At finite length, finite size corrections in $1/N$ carry a
different numerical prefactor.

\bibitem{Leggett:1987}
A.\ J.\ Leggett, S.\ Chakravarty, A.\ T.\ Dorsey, M.\ P.\ A.\ Fisher, A.\ Garg, and W.\ Zwerger,
Rev.\ Mod.\ Phys.\ {\bf 59}, 1  (1987).


\end{thebibliography}
\end{document}